\documentstyle[12pt,epsf,rotating,axodraw]{article}

\catcode`@=11
\def\citer{\@ifnextchar [{\@tempswatrue\@citexr}{\@tempswafalse\@citexr[]}}
 
\def\@citexr[#1]#2{\if@filesw\immediate\write\@auxout{\string\citation{#2}}\fi
  \def\@citea{}\@cite{\@for\@citeb:=#2\do
    {\@citea\def\@citea{--\penalty\@m}\@ifundefined
       {b@\@citeb}{{\bf ?}\@warning
       {Citation `\@citeb' on page \thepage \space undefined}}%
\hbox{\csname b@\@citeb\endcsname}}}{#1}}
\catcode`@=12
 
\newcommand{\lsim}{\raisebox{-0.13cm}{~\shortstack{$<$ \\[-0.07cm] $\sim$}}~}
\newcommand{\gsim}{\raisebox{-0.13cm}{~\shortstack{$>$ \\[-0.07cm] $\sim$}}~}

\newcommand{\beq}{\begin{equation}}
\newcommand{\eeq}{\end{equation}}
\newcommand{\bea}{\begin{eqnarray}}
\newcommand{\eea}{\end{eqnarray}}
\newcommand{\CP}{\mbox{${\cal CP}$}}

\newcommand{\MS}{\mbox{$\overline{\mathrm{MS}}$~}}

\newcommand{\tgb}{\mbox{tg$\beta$}}

\def\msb{\overline{\rm MS}}

\newbox\mycount
\newcommand{\ctowidth}[2]{ \setbox\mycount=\hbox{$#2$}
                          \hbox to \wd\mycount{$ \hss #1 \hss $} }
\newcommand{\ltowidth}[2]{ \setbox\mycount=\hbox{$#2$}
                          \hbox to \wd\mycount{$\hskip0pt plus0pt minus1fil
                           #1 \hfill $} }
\newcommand{\rtowidth}[2]{ \setbox\mycount=\hbox{$#2$}
                          \hbox to \wd\mycount{$\hfill #1
                          \hskip0pt plus0pt minus1fil$} }

\def\draftdate{\relax}
\def\mda{\relax}
\def\mua{\relax}
\def\mla{\relax}
\def\Mda{\relax}
\def\Mua{\relax}
\def\Mla{\relax}
\def\draft{
\def\thtystars{******************************}
\def\sixtystars{\thtystars\thtystars}
\typeout{}
\typeout{\sixtystars**}
\typeout{* Draft mode!
         For final version remove \protect\draft\space in source file *}
\typeout{\sixtystars**}
\typeout{}
\def\draftdate{\today}
\def\mua{\marginpar[\boldmath\hfil$\uparrow$]%
                   {\boldmath$\uparrow$\hfil}%
                    \typeout{marginpar: $\uparrow$}\ignorespaces}
\def\mda{\marginpar[\boldmath\hfil$\downarrow$]%
                   {\boldmath$\downarrow$\hfil}%
                    \typeout{marginpar: $\downarrow$}\ignorespaces}
\def\mla{\marginpar[\boldmath\hfil$\rightarrow$]%
                   {\boldmath$\leftarrow $\hfil}%
                    \typeout{marginpar: $\leftrightarrow$}\ignorespaces}
\def\Mua{\marginpar[\boldmath\hfil$\Uparrow$]%
                   {\boldmath$\Uparrow$\hfil}%
                    \typeout{marginpar: $\Uparrow$}\ignorespaces}
\def\Mda{\marginpar[\boldmath\hfil$\Downarrow$]%
                   {\boldmath$\Downarrow$\hfil}%
                    \typeout{marginpar: $\Downarrow$}\ignorespaces}
\def\Mla{\marginpar[\boldmath\hfil$\Rightarrow$]%
                   {\boldmath$\Leftarrow $\hfil}%
                    \typeout{marginpar: $\Leftrightarrow$}\ignorespaces}
\overfullrule 5pt
\oddsidemargin -15mm
\marginparwidth 29mm
}

\begin{document}

\renewcommand{\thefootnote}{\fnsymbol{footnote}}

\begin{titlepage}

\begin{flushright}
DESY 98-028 \\
BNL-HET-98/15 \\
CERN-TH/98-135 \\
hep-ph/9805244 \\
May 1998
\end{flushright}

\vspace*{1cm}

\begin{center}

{\Large\sc Neutral Higgs-Boson Pair Production \\[0.5cm] 
at Hadron Colliders: QCD Corrections}

\vspace{1cm}

{\sc S.\ Dawson$^1$, S.\ Dittmaier$^2$ and M.\ Spira\footnote{Heisenberg
Fellow.}$^3$} \\
\vspace{1cm}

$^1$ {\it Physics Department\footnote{
Supported by U.S. Department of Energy contract number
DE-AC02-98CH10886.}, Brookhaven National Laboratory, Upton, NY
11973, USA} \\[0.5cm]

$^2$ {\it Theory Division, CERN, CH--1211 Geneva 23, Switzerland}
\\[0.5cm]

$^3$ {\it II.\ Institut f\"ur Theoretische Physik\footnote{Supported by
Bundesministerium f\"ur Bildung und Forschung (BMBF), Bonn, Germany, under
Contract 05~7~HH~92P~(5), and by EU Program {\it Human Capital and Mobility}
through Network {\it Physics at High Energy Colliders} under Contract
CHRX--CT93--0357 (DG12 COMA).}, Universit\"at Hamburg, Luruper Chaussee 149,
D--22761 Hamburg, Germany}

\end{center}

\vspace{1cm}

\begin{abstract}
\normalsize
\noindent
Neutral Higgs-boson pair production provides the possibility of studying the
trilinear Higgs couplings at future high-energy colliders. We present the QCD
corrections to the gluon-initiated processes in the limit of a
heavy top quark in the loops and the Drell--Yan-like pair production of
scalar and pseudoscalar Higgs particles. The $pp$ cross sections 
are discussed for LHC energies within the Standard Model
and its minimal supersymmetric extension. The QCD corrections are large,
enhancing the total cross sections significantly.
\end{abstract}

\end{titlepage}

\renewcommand{\thefootnote}{\arabic{footnote}}

\setcounter{footnote}{0}
\setcounter{page}{2}

\section[]{Introduction}
The Higgs mechanism is a cornerstone of the Standard Model (SM) and
its supersymmetric extensions.
Thus, the search for Higgs bosons is one of  the most important 
endeavours 
at future high-energy experiments.  In the SM one Higgs doublet, $\Phi$,
has to be introduced in order to break the electroweak symmetry, leading to
the existence of one elementary Higgs boson, $H$ \cite{higgs}.
The scalar sector of the SM is uniquely fixed by the vacuum expectation
value $v$ of the Higgs doublet and the mass $M_H$ of the physical Higgs
boson \cite{hunter}. Once a Higgs particle is found, it is necessary
to investigate its properties in order to reconstruct the Higgs potential
and to verify that it is indeed the SM Higgs boson. A
first step in  this direction is the measurement of the trilinear
self-couplings,
which are uniquely specified by the scalar potential
\begin{equation}
V=\lambda\biggl(\Phi^\dagger \Phi -{v^2\over 2}\biggr)^2.
\end{equation}
The parameter $\lambda$ defines the strength
of the Higgs self-interactions. In the SM it is given by
$\lambda=M_H^2/(2v^2)$.
At tree level, $\lambda$ can only be probed through multiple 
Higgs-boson interactions and there are, at present, 
no direct experimental limits on $\lambda$. In extensions of the SM, 
such as models with an extended scalar sector, with composite particles 
or with supersymmetric partners,
the self-couplings of the Higgs boson may be significantly different from
the SM predictions. 
The limits that may be obtained on the trilinear
self-coupling of the Higgs boson at the LHC and the impact of QCD
corrections represent a particular topic of this paper.   

Since the minimal supersymmetric extension of the Standard Model (MSSM)
requires the introduction of two Higgs doublets in order to preserve
supersymmetry, there are five elementary Higgs particles, two
\CP-even ($h,H$)%
\footnote{We have taken care that no confusion can arise
from using the same symbol $H$ for the SM and the heavy \CP-even MSSM Higgs
particle.},
one \CP-odd ($A$) and two charged ones ($H^\pm$). This leads to a large
variety of self-interactions among them. At lowest order all couplings and
masses of the MSSM Higgs sector are fixed by two independent input parameters,
which are generally chosen as $\tgb=v_2/v_1$, the ratio of the two vacuum
expectation values $v_{1,2}$, and the pseudoscalar 
Higgs-boson mass $M_A$.
The self-interactions 
among the Higgs bosons (at lowest order) are
given in terms of the electroweak gauge couplings, $\tgb$ and $M_A$,
and may be quite different from the 
ones of the SM, which are governed by the parameter $\lambda$.
Higher-order corrections to the MSSM Higgs sector turn out to be important 
owing to the large top-quark mass $m_t$ \citer{mssmrad1,mssmrad2}.
They increase the upper bound on
the light scalar Higgs mass $M_h$ from the $Z$-boson mass $M_Z$
to about 130~GeV, along with
altering the Higgs-boson self-couplings with contributions proportional
to $G_F m_t^4/M_Z^2$.

Higgs-boson 
pairs can be produced by several mechanisms at hadron colliders:
\begin{itemize}  
\item 
 Higgs-strahlung $W^*/Z^* \to \phi_1 \phi_2 W/Z$ \cite{ppvhh},
\item  
 vector-boson fusion $WW,ZZ \to \phi_1 \phi_2$ \cite{ppvvhh},
\item   
 associated production $Z^*\to Ah,AH$,
\item  
 Higgs radiation off top and bottom quarks $gg,q\bar q
 \to Q\bar Q \phi_1 \phi_2$ \cite{ppqqhh},
\item  
 gluon--gluon collisions $gg\to \phi_1\phi_2$ \cite{gghhsm,gghhlo}.
\end{itemize}
At the LHC, gluon fusion is the dominant source of Higgs-boson pairs, 
although in some regions of the MSSM parameter space, vector-boson 
fusion \cite{djlhc} can be important.
Note, however, that $gg\to HA$ represents an exceptional case, since
this channel is suppressed with respect to
the Drell--Yan-like process $q\bar q\to Z^* \to HA$, so that it
will be very difficult to separate the 
gluon-fusion process in this case experimentally. The gluon-fusion process
$gg\to hA$, on the other hand, is competitive with the Drell--Yan-like process
$q\bar q\to Z^* \to hA$.

In this paper we present the QCD corrections to the Drell--Yan-like production
and the gluon--gluon collision processes. 
The gluon-fusion processes
are, in the SM, mediated by triangle and box loops of
top and bottom quarks;
in the SM, the contributions of the bottom quark can always be
neglected. In the MSSM, the squark 
contributions will be suppressed if the squarks are heavier than 
$\sim 400$~GeV, and,
for small $\tgb$, the top-quark loops dominate the $gg$ cross sections. 
The QCD corrections to the 
gluon-fusion processes have been obtained in the 
limit of a heavy top quark by means of low-energy theorems and 
also by explicitly expanding all relevant one- 
and two-loop diagrams. 
The results are expected to be valid for small $\tgb$ in the
MSSM and below the $t\bar t$ threshold of the top-quark loops in both 
the MSSM and the SM, since in this regime 
effects of a finite top-quark mass are expected to be small.
In the case of 
single-Higgs production the same procedure reproduces the
known exact result for the NLO cross section within 5\%,
for Higgs-boson masses below $2m_t$ \cite{soft}.
The considered QCD corrections are important 
in the process of extracting
limits on the Higgs-boson self-couplings reliably.
 
The paper is organized as follows. In the next section the low-energy theorems
for the interactions of gluons with light Higgs bosons will be reviewed,
and the relevant interactions in the heavy-quark limit will be constructed. 
In Section~\ref{se:qcdcorr} the details of the calculation will be described,
and in Section~\ref{se:results} we present
the results for the SM and MSSM Higgs bosons. In Section~\ref{se:concl}
we give some conclusions.

\section{Low-Energy Theorems}
In the low-energy limit of vanishing Higgs four-momentum, the Higgs-field
operator acts as a constant field.
In this limit it is possible to derive an effective Lagrangian for
the interactions of the Higgs bosons with gauge bosons,
which is valid for light Higgs bosons. This effective Lagrangian has been
successfully used to compute the QCD corrections to a number of
processes, in particular to
single-Higgs production from gluon fusion at the
LHC \cite{gghlim,higgsqcd}. In this case, the 
result of using the low-energy 
theorems has been shown to agree with the exact two-loop
calculation to better than $10\%$ even for $M_H$ as large as $1$ TeV.
This lends legitimacy to our use of the low-energy theorems to
compute QCD corrections to multiple 
Higgs-boson production via gluon fusion.  

In the limit of vanishing Higgs four-momentum
the entire interaction of the scalar Higgs particles $H_i$ with 
a heavy quark $Q$ can be generated by the substitution
\begin{equation}
m_Q^0 \to m_Q^0 + \eta_Q^0 \sum_i g^i_Q H_i
\label{eq:subst}
\end{equation}
in the Lagrangian of a heavy quark of bare mass $m_Q^0$ \cite{let0}, 
where $\eta_Q^0=m_Q^0/v$ denotes the bare SM Yukawa coupling,
which must not be included in the substitution. The symbol $g^i_Q$
is the relative strength of the heavy-quark Yukawa coupling, 
\begin{equation}
{\cal L}_{\mathrm{Yuk}} = -\eta_Q\sum_i g^i_Q {\overline Q} Q H_i.
\label{eq:yukdef}
\end{equation} 
In the SM, we have $g^i_Q=1$. The expressions for $g^i_Q$ in the 
MSSM are given in Ref. \cite{higgsqcd}. 
At higher orders this substitution
has to be performed for the 
unrenormalized parameters \cite{higgsqcd,let1}. 
In the following we restrict our analysis to the top-quark contributions.
At next-to-leading order (NLO) the effective
interaction between several scalar Higgs fields and gluons can be obtained
from the radiatively corrected effective Lagrangian of gluon fields,
\begin{equation}
{\cal L}_{gg} = -\frac{1}{4} G^{a\mu\nu} G^a_{\mu\nu} \left[ 1+\Pi_{gg}^t(0)
\right],
\end{equation}
with $\Pi_{gg}^t(0)$ denoting the top-quark contribution to the 
unrenormalized gluon vacuum polarization at zero-momentum transfer. 
At two-loop order, we have
\begin{equation}
\Pi_{gg}^t(0) = \frac{\alpha_s^{(5)}}{\pi} \Gamma(1+\epsilon) \left(
\frac{4\pi\mu^2}{(m_t^0)^2} \right)^{\epsilon} \left\{ \frac{1}{6\epsilon}
+ \frac{\alpha_s^{(5)}}{\pi} \Gamma(1+\epsilon) \left(
\frac{4\pi\mu^2}{(m_t^0)^2} \right)^{\epsilon} \left[ \frac{1}{16\epsilon}
+ {\cal O}(\epsilon^0) \right] + {\cal O}(\alpha_s^2) \right\},
\label{eq:pitgg}
\end{equation}
where the strong coupling constant $\alpha_s^{(5)}$ includes 
five light flavours. 
This means that the top-quark contribution 
to the running of $\alpha_s$ has been subtracted at vanishing
momentum transfer. 
Hereafter, we drop the superscript $5$ on $\alpha_s$.
Performing the substitution Eq.\ (\ref{eq:subst})
and renormalizing the bare top mass $m_t^0$ via
\begin{equation}
m_t^0 = m_t \left[ 1 - \frac{\alpha_s}{\pi} \Gamma(1+\epsilon)
\left(\frac{4\pi\mu^2}{m_t^2}\right)^\epsilon
\left( \frac{1}{\epsilon} + \frac{4}{3} \right) + {\cal O}(\alpha_s^2)
\right],
\end{equation}
where $m_t$ denotes the 
pole mass, we end up with the NLO result
\begin{equation}
{\cal L}_{H^ngg} = \frac{\alpha_s}{12\pi} G^{a\mu\nu} G^a_{\mu\nu} \log
\left[ 1+\sum_i g^i_t \frac{H_i}{v} \right] \left\{ 1+\frac{11}{4}
\frac{\alpha_s}{\pi} \right\}.
\label{eq:hgg}
\end{equation}
  
The interaction of even numbers of pseudoscalar Higgs bosons with gluons can
be determined analogously from Eq.\ (\ref{eq:pitgg}) by substituting
\cite{let1}
\begin{equation}
(m_t^0)^2 \to (m_t^0)^2 + (g_t^A \eta_t^0 A)^2,
\end{equation}
leading to
\begin{equation}
{\cal L}_{A^{2n}gg} = \frac{\alpha_s}{24\pi} G^{a\mu\nu} G^a_{\mu\nu} \log
\left[ 1+\left(g^A_t \frac{A}{v}\right)^2 \right] \left\{ 1+\frac{11}{4}
\frac{\alpha_s}{\pi} \right\} .
\end{equation}
The case of odd numbers of pseudoscalar Higgs bosons can be derived from the
ABJ anomaly \cite{abj} in the divergence of the axial vector current
\cite{higgsqcd,let1,agg}. 
The interactions that are relevant in our case are\footnote{Note that in the
earlier Refs.\ \cite{higgsqcd,let1,agg} a factor of 1/2 is missing in the
effective Lagrangians for the single pseudoscalar Higgs-boson coupling to
gluons.}
\begin{eqnarray}
{\cal L}_{Agg} & = & g^A_t \frac{\alpha_s}{8\pi} G^{a\mu\nu} \tilde
G^a_{\mu\nu} \frac{A}{v}, \nonumber \\
{\cal L}_{AHgg} & = & -g^A_t g^H_t \frac{\alpha_s}{8\pi} G^{a\mu\nu} \tilde
G^a_{\mu\nu} \frac{AH}{v^2},
\label{eq:agg}
\end{eqnarray}
where $\tilde G^a_{\mu\nu} = \frac{1}{2} \epsilon_{\mu\nu\alpha\beta}
G^{a\alpha\beta}$ denotes the dual gluon 
field-strength tensor. 
Owing to the
Adler--Bardeen theorem \cite{abt} there are no higher-order corrections to the
effective Lagrangians involving odd numbers of pseudoscalar Higgs bosons.
  
Figure~\ref{fg:feyneff} summarizes the Feynman rules for the effective 
interactions between two gluons and one or two Higgs bosons; the rules
can be read off from
Eqs.\ (\ref{eq:hgg})--(\ref{eq:agg}).
\begin{figure}[p]
\begin{picture}(440,100)(-50,0)
\Gluon(0,80)(50,50){3}{6}
\Gluon(0,20)(50,50){3}{6}
\DashLine(50,50)(100,50){5}
\GCirc(50,50){10}{0.5}
\put(105,46){$H$}
\put(-40,18){$g^a_\mu(k_1)$}
\put(-40,78){$g^b_\nu(k_2)$}
\put(140,46){$\displaystyle i \delta_{ab} \frac{\alpha_s}{3\pi v} 
g_t^H \left\{ -g^{\mu\nu} (k_1\cdot k_2) + k_1^\nu k_2^\mu \right\} 
\left(1+\frac{11}{4}\frac{\alpha_s}{\pi} \right)$}
\end{picture} 
\\
\begin{picture}(440,100)(-50,0)
\Gluon(0,80)(50,50){3}{6}
\Gluon(0,20)(50,50){3}{6}
\DashLine(50,50)(100,50){5}
\GCirc(50,50){10}{0.5}
\put(105,46){$A$}
\put(-40,18){$g^a_\mu(k_1)$}
\put(-40,78){$g^b_\nu(k_2)$}
\put(140,46){$\displaystyle
i \delta_{ab} \frac{\alpha_s}{2\pi v} g_t^A \epsilon^{\mu\nu\alpha\beta}
k_{1\alpha} k_{2\beta}$}
\end{picture} 
\\
\begin{picture}(440,100)(-50,0)
\Gluon(0,80)(50,50){3}{6}
\Gluon(0,20)(50,50){3}{6}
\DashLine(50,50)(100,80){5}
\DashLine(50,50)(100,20){5}
\GCirc(50,50){10}{0.5}
\put(105,76){$H_1$}
\put(105,16){$H_2$}
\put(-40,18){$g^a_\mu(k_1)$}
\put(-40,78){$g^b_\nu(k_2)$}
\put(140,46){$\displaystyle -i \delta_{ab} \frac{\alpha_s}{3\pi v^2}
g_t^{H_1} g_t^{H_2} \left\{ - g^{\mu\nu} (k_1\cdot k_2) + k_1^\nu k_2^\mu 
\right\} \left(1+\frac{11}{4}\frac{\alpha_s}{\pi} \right)$}
\end{picture} 
\\
\begin{picture}(440,100)(-50,0)
\Gluon(0,80)(50,50){3}{6}
\Gluon(0,20)(50,50){3}{6}
\DashLine(50,50)(100,80){5}
\DashLine(50,50)(100,20){5}
\GCirc(50,50){10}{0.5}
\put(105,76){$A$}
\put(105,16){$A$}
\put(-40,18){$g^a_\mu(k_1)$}
\put(-40,78){$g^b_\nu(k_2)$}
\put(140,46){$\displaystyle i \delta_{ab} \frac{\alpha_s}{3\pi v^2}
(g_t^A)^2 \left\{ - g^{\mu\nu} (k_1\cdot k_2) + k_1^\nu k_2^\mu \right\} 
\left(1+\frac{11}{4}\frac{\alpha_s}{\pi} \right)$}
\end{picture} 
\\
\begin{picture}(440,100)(-50,0)
\Gluon(0,80)(50,50){3}{6}
\Gluon(0,20)(50,50){3}{6}
\DashLine(50,50)(100,80){5}
\DashLine(50,50)(100,20){5}
\GCirc(50,50){10}{0.5}
\put(105,76){$H$}
\put(105,16){$A$}
\put(-40,18){$g^a_\mu(k_1)$}
\put(-40,78){$g^b_\nu(k_2)$}
\put(140,46){$\displaystyle -i \delta_{ab} \frac{\alpha_s}{2\pi v^2}
g_t^H g_t^A \epsilon^{\mu\nu\alpha\beta} k_{1\alpha} k_{2\beta}$}
\end{picture}
\caption[]{\it \label{fg:feyneff} Feynman rules for the effective
interactions of Higgs bosons with gluons in the heavy-quark limit,
including NLO corrections.}
\end{figure}
These Feynman rules can now be used to compute Higgs interactions
beyond the lowest order.
We recall that there is no contribution of light quarks (which are
considered to be massless) to the effective couplings, but note that 
light-quark loops have to be included when the Higgs bosons do not 
directly couple to the quark loops. Such contributions arise, in 
particular, in 
$gg\to Z^* \to hA,HA$ and cannot be obtained from the 
low-energy theorems.

\section{QCD Corrections}
\label{se:qcdcorr}

\subsection{Gluon fusion: basic definitions}

\begin{figure}[t]
\begin{picture}(100,100)(-30,-5)
\SetScale{0.8}
\Gluon(0,100)(50,100){3}{6}
\Gluon(0,0)(50,0){3}{6}
\ArrowLine(50,100)(100,50)
\ArrowLine(100,50)(50,0)
\ArrowLine(50,0)(50,100)
\DashLine(100,50)(150,50){5}
\DashLine(150,50)(200,100){5}
\DashLine(150,50)(200,0){5}
\put(165,76){$\phi_1$}
\put(165,-5){$\phi_2$}
\put(90,50){$\phi,Z$}
\put(-15,-3){$g$}
\put(-15,78){$g$}
\put(20,38){$t,b$}
\SetScale{1}
\end{picture}
\begin{picture}(100,100)(-150,-5)
\SetScale{0.8}
\Gluon(0,100)(50,100){3}{6}
\Gluon(0,0)(50,0){3}{6}
\ArrowLine(50,100)(140,100)
\ArrowLine(140,100)(140,0)
\ArrowLine(140,0)(50,0)
\ArrowLine(50,0)(50,100)
\DashLine(140,100)(190,100){5}
\DashLine(140,0)(190,0){5}
\put(165,76){$\phi_1$}
\put(165,-5){$\phi_2$}
\put(-15,-3){$g$}
\put(-15,78){$g$}
\put(20,38){$t,b$}
\SetScale{1}
\end{picture}
\caption[]{\it \label{fg:lodia} Generic diagrams describing neutral 
Higgs-boson pair production in gluon--gluon collisions ($\phi,\phi_i = h,H,A$).}
\end{figure}
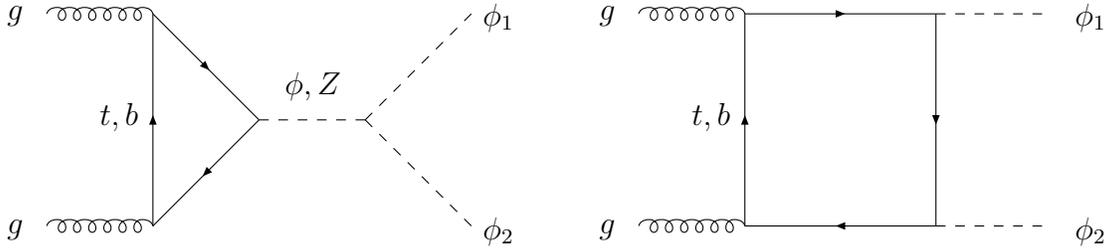

At leading order (LO) 
neutral-Higgs pair production via gluon fusion is mediated by 
triangle and box diagrams of heavy quarks, 
as exemplified in Fig.~\ref{fg:lodia}.
In the heavy-quark limit, the fermion triangles and boxes can be replaced
by the effective vertices of Fig.~\ref{fg:feyneff}.
Throughout this analysis,
we choose the squark masses to be 1~TeV so that
squark-loop contributions can be neglected in the MSSM case.
Generically the 
partonic LO cross section can be expressed as \cite{gghhlo}
\begin{equation}
\hat \sigma_{\mathrm{LO}}(gg\to\phi_1\phi_2)
 = \int_{\hat t_-}^{\hat t_+} d\hat t \,
\frac{G_F^2 \alpha_s^2(\mu)}{256 (2\pi)^3} \left\{ \left|
C_\triangle F_\triangle + C_\Box F_\Box \right|^2 + \left| C_\Box G_\Box
\right|^2 \right\}.
\label{eq:gghhlo}
\end{equation}
The Mandelstam variables for the parton process are given by
\begin{eqnarray}
\hat s &=& Q^2, 
\nonumber \\
\hat t &=& -\frac{1}{2} 
\left[Q^2-m_1^2-m_2^2-\sqrt{\lambda(Q^2, m_1^2, m_2^2)}\cos\theta\right], 
\nonumber \\
\hat u &=& -\frac{1}{2} 
\left[Q^2-m_1^2-m_2^2+\sqrt{\lambda(Q^2, m_1^2, m_2^2)}\cos\theta\right], 
\end{eqnarray}
where $\theta$ is the scattering angle in
the partonic c.m.\ system with invariant mass $Q$, and
\begin{equation}
\lambda(x,y,z) = (x-y-z)^2 - 4 yz.
\end{equation}
The integration limits
\begin{equation}
\hat t_\pm = -\frac{1}{2} \left[ Q^2 - m_1^2 - m_2^2 \mp
\sqrt{\lambda(Q^2, m_1^2, m_2^2)} \right]
\end{equation}
in Eq.\ (\ref{eq:gghhlo})
correspond to $\cos\theta=\pm 1$.
The scale parameter $\mu$ is 
the renormalization scale.
The complete dependence on the fermion masses
is contained in the functions $F_\triangle$, $F_\Box$, and $G_\Box$.  
The full expressions of the form factors $F_\triangle$, $F_\Box$, $G_\Box$,
including the exact dependence on the fermion masses, can
be found in Ref.~\cite{gghhlo}.

The couplings $C_\triangle$ and $C_\Box$ and the form factors $F_\triangle,
F_\Box, G_\Box$ in the heavy-quark limit are given by: 
\renewcommand{\labelenumi}{(\roman{enumi})}
\begin{enumerate}
\item \underline{SM:}
\begin{eqnarray}
C_\triangle & = & \lambda_{HHH} \frac{M_Z^2}{\hat s - M_H^2 + iM_H \Gamma_H}, 
\qquad
C_\Box = 1, \nonumber \\ 
F_\triangle & \to & \frac{2}{3}, \qquad 
F_\Box \to -\frac{2}{3}, \nonumber\\
G_\Box & \to & 0, 
\end{eqnarray}
with the trilinear coupling $\lambda_{HHH} = 3 M_H^2/M_Z^2$. 

\item \underline{MSSM:} \\[0.3cm]
The couplings for the processes $gg\to \phi_1 \phi_2$ are generically defined
as ($\phi,\phi_i = h,H,A$)
\begin{equation}
C_\triangle^{\phi} = \lambda_{\phi_1 \phi_2 \phi}
\frac{M_Z^2}{\hat s - M_\phi^2 + iM_\phi \Gamma_\phi} g_t^\phi, \qquad
C_\Box = g_t^{\phi_1} g_t^{\phi_2},
\end{equation}
where $\phi$ denotes the Higgs 
particles of the $s$-channel contributions. The
trilinear couplings $\lambda_{\phi_1\phi_2\phi}$ and the normalized Yukawa
couplings $g_t^\phi$ can be found in Ref.\ \cite{gghhlo}. The individual
expressions in the heavy-quark limit can be summarized as:
\begin{description}
\item[$\phi_1\phi_2 = hh,hH,HH$:]
\begin{eqnarray}
C_\triangle & = & C_\triangle^h + C_\triangle^H, \nonumber\\ 
F_\triangle & \to & \frac{2}{3}, \qquad
F_\Box \to -\frac{2}{3}, \nonumber\\
G_\Box & \to & 0.
\end{eqnarray}

\item[$\phi_1\phi_2 = hA,HA$:]
\begin{eqnarray}
C_\triangle & = & C_\triangle^A + C_\triangle^Z, \qquad
C_\triangle^Z = \lambda_{ZAh,ZAH} \,
\frac{M_Z^2}{\hat s - M_Z^2 + iM_Z \Gamma_Z} a_t, \nonumber \\ 
F_\triangle^A & \to & 1, \qquad
F_\triangle^Z \to \frac{\hat s - M_Z^2}{M_Z^2 \hat s} (M_{h,H}^2 - M_A^2),
\qquad
F_\Box \to -1, \nonumber \\
G_\Box & \to & 0,
\end{eqnarray}
where $a_t = 1$ denotes the axial charge of the top quark.

\item[$\phi_1\phi_2 = AA$:]
\begin{eqnarray}
C_\triangle & = & C_\triangle^h + C_\triangle^H, \nonumber \\
F_\triangle & \to & \frac{2}{3}, \qquad 
F_\Box \to \frac{2}{3}, \nonumber \\
G_\Box & \to & 0.
\end{eqnarray}
\end{description}
\end{enumerate}
It should be noted that 
owing to the Ward identities for the $Zgg$ vertex
only the pseudoscalar Goldstone component of the $Z$~bosons
contributes to $F_\triangle^Z$ in the case of 
$hA$ and $HA$ production in the heavy-quark limit. 
The anomaly contributions of the top and bottom
quarks cancel.

The QCD corrections consist of two-loop virtual corrections, generated by
gluon exchange between the 
quark lines and/or external gluons, and one-loop
real corrections with an additional gluon or light quark in the final state. We
have evaluated the QCD corrections in the 
heavy-quark limit by means of two different methods: 
(i) using the effective couplings based on the low-energy theorems, 
as presented in the previous section, and 
(ii) explicitly expanding all relevant one- 
and two-loop diagrams in the inverse heavy-quark 
mass. In the following we shall describe the details of both approaches.

\subsection{Low-energy theorems}
\begin{figure}[t]
\begin{picture}(100,90)(-30,-5)
\SetScale{0.8}
\Gluon(0,100)(25,75){3}{3}
\Gluon(25,75)(50,50){3}{3}
\Gluon(0,0)(25,25){3}{3}
\Gluon(25,25)(50,50){3}{3}
\GlueArc(50,50)(35.355,135,225){3}{5}
\DashLine(50,50)(100,100){5}
\DashLine(50,50)(100,0){5}
\GCirc(50,50){10}{0.5}
\put(85,76){$\phi_1$}
\put(85,-5){$\phi_2$}
\put(-15,-3){$g$}
\put(-15,78){$g$}
\put(-30,38){$(a)$}
\SetScale{1}
\end{picture}
\begin{picture}(100,90)(-60,-5)
\SetScale{0.8}
\Gluon(0,100)(25,75){3}{3}
\Gluon(25,75)(50,50){3}{3}
\Gluon(0,0)(25,25){3}{3}
\Gluon(25,25)(50,50){3}{3}
\GlueArc(50,50)(35.355,135,225){3}{5}
\DashLine(50,50)(90,50){5}
\DashLine(90,50)(140,100){5}
\DashLine(90,50)(140,0){5}
\GCirc(50,50){10}{0.5}
\put(115,76){$\phi_1$}
\put(115,-5){$\phi_2$}
\put(50,50){$\phi,Z$}
\put(-15,-3){$g$}
\put(-15,78){$g$}
\SetScale{1}
\end{picture}
\begin{picture}(100,90)(-115,-5)
\SetScale{0.8}
\Gluon(0,100)(50,100){3}{4}
\Gluon(0,0)(50,0){3}{4}
\Gluon(50,0)(50,100){3}{10}
\DashLine(50,100)(100,100){5}
\DashLine(50,0)(100,0){5}
\GCirc(50,100){10}{0.5}
\GCirc(50,0){10}{0.5}
\put(85,76){$\phi_1$}
\put(85,-5){$\phi_2$}
\put(-15,-3){$g$}
\put(-15,78){$g$}
\SetScale{1}
\end{picture} \\
\begin{picture}(100,110)(-30,-5)
\SetScale{0.8}
\Gluon(0,100)(25,75){3}{3}
\Gluon(25,75)(50,50){3}{3}
\Gluon(25,75)(50,100){3}{3}
\Gluon(0,0)(50,50){3}{6}
\DashLine(50,50)(100,100){5}
\DashLine(50,50)(100,0){5}
\GCirc(50,50){10}{0.5}
\put(85,76){$\phi_1$}
\put(85,-5){$\phi_2$}
\put(-15,-3){$g$}
\put(-15,78){$g$}
\put(45,78){$g$}
\put(-30,38){$(b)$}
\SetScale{1}
\end{picture}
\begin{picture}(100,110)(-60,-5)
\SetScale{0.8}
\Gluon(0,100)(50,50){3}{6}
\Gluon(0,0)(50,50){3}{6}
\Gluon(50,50)(100,100){3}{6}
\DashLine(50,50)(100,50){5}
\DashLine(50,50)(100,0){5}
\GCirc(50,50){10}{0.5}
\put(85,36){$\phi_1$}
\put(85,-5){$\phi_2$}
\put(-15,-3){$g$}
\put(-15,78){$g$}
\put(85,78){$g$}
\SetScale{1}
\end{picture}
\begin{picture}(100,110)(-115,-5)
\SetScale{0.8}
\ArrowLine(0,100)(50,100)
\ArrowLine(50,100)(100,100)
\Gluon(50,100)(50,50){3}{4}
\Gluon(0,0)(50,50){3}{6}
\DashLine(50,50)(100,50){5}
\DashLine(50,50)(100,0){5}
\GCirc(50,50){10}{0.5}
\put(85,36){$\phi_1$}
\put(85,-5){$\phi_2$}
\put(-15,-3){$g$}
\put(-15,78){$q$}
\put(85,78){$q$}
\SetScale{1}
\end{picture}
\caption[]{\it \label{fg:gghhdiaeff} Typical effective diagrams contributing
to the (a) virtual and (b) real corrections to neutral Higgs-boson pair 
production.}
\end{figure}
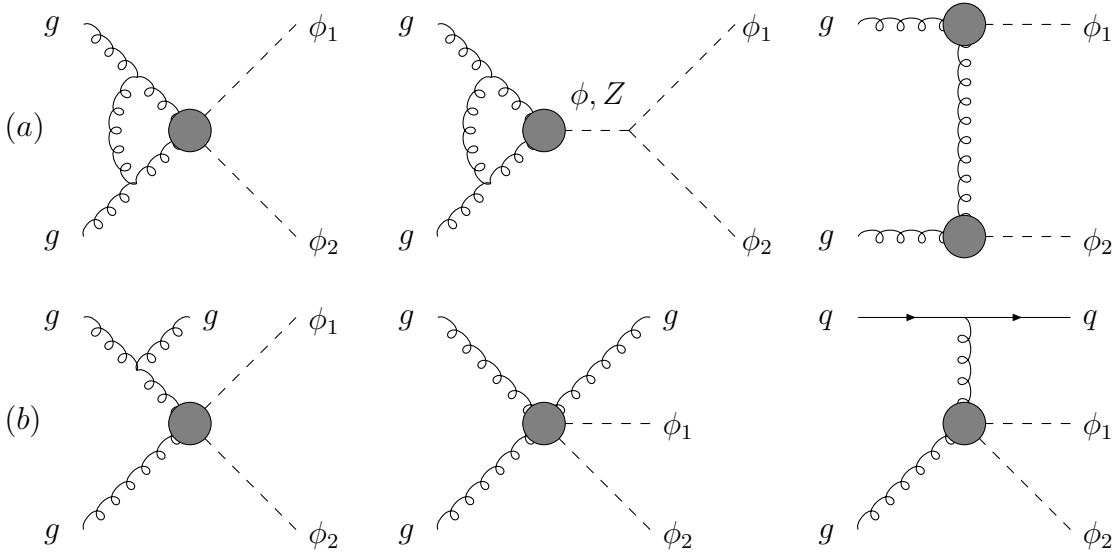

We are now in a position to compute the NLO
corrections to Higgs-boson pair production.  
Typical effective diagrams contributing to the virtual and real
corrections  are presented in Fig.~\ref{fg:gghhdiaeff}.
Adopting the Feynman rules of Fig.~\ref{fg:feyneff} for the effective
interactions, the calculation has been carried out 
in dimensional regularization with $n=4-2\epsilon$ dimensions. The strong
coupling has been renormalized in the \MS scheme including 
five light-quark
flavours, i.e.\ decoupling the top quark 
in the running of $\alpha_s$. After summing the
virtual and real corrections the infrared singularities cancel. However,
collinear initial-state singularities are left over in the partonic cross
sections. 
Those divergences have been absorbed into the NLO parton densities,
defined in the \MS scheme with 
five light-quark flavours. We end up with
finite results, which can be cast into the form
\begin{equation}
\sigma_{\mathrm{NLO}}(pp \rightarrow \phi_1 \phi_2 + X) = 
\sigma_{\mathrm{LO}} + \Delta
\sigma_{\mathrm{virt}} + \Delta\sigma_{gg} + \Delta\sigma_{gq} + \Delta\sigma_{q\bar{q}},
\label{eq:gghqcd}
\end{equation}
with the individual contributions
\begin{eqnarray}
\sigma_{\mathrm{LO}} & = & \int_{\tau_0}^1 d\tau~\frac{d{\cal L}^{gg}}{d\tau}~
\hat\sigma_{\mathrm{LO}}(Q^2 = \tau s), 
\nonumber \\ 
\Delta \sigma_{\mathrm{virt}} & = & \frac{\alpha_s(\mu)} {\pi}\int_{\tau_0}^1 d\tau~
\frac{d{\cal L}^{gg}}{d\tau}~\hat \sigma_{\mathrm{LO}}(Q^2=\tau s)~C, 
\nonumber \\ 
\Delta \sigma_{gg} & = & \frac{\alpha_{s}(\mu)} {\pi} \int_{\tau_0}^1 d\tau~
\frac{d{\cal L}^{gg}}{d\tau} \int_{\tau_0/\tau}^1 \frac{dz}{z}~
\hat\sigma_{\mathrm{LO}}(Q^2 = z \tau s)
\left\{ - z P_{gg} (z) \log \frac{M^{2}}{\tau s} \right. 
\nonumber \\
& & \left. \hspace*{3.0cm} {} - \frac{11}{2} (1-z)^3 + 6 [1+z^4+(1-z)^4]
\left(\frac{\log (1-z)}{1-z} \right)_+ \right\}, 
\nonumber \\ 
\Delta \sigma_{gq} & = & \frac{\alpha_{s}(\mu)} {\pi} \int_{\tau_0}^1 d\tau
\sum_{q,\bar{q}} \frac{d{\cal L}^{gq}}{d\tau} \int_{\tau_0/\tau}^1 \frac{dz}{z}~
\hat \sigma_{\mathrm{LO}}(Q^2 = z \tau s)
\left\{ -\frac{z}{2} P_{gq}(z) \log\frac{M^{2}}{\tau s(1-z)^2} \right. 
\nonumber \\
& & \left. \hspace*{9.0cm} {} + \frac{2}{3}z^2 - (1-z)^2 
\vphantom{\frac{M^{2}}{\tau s(1-z)^2}} \right\},
\nonumber \\ 
\Delta \sigma_{q\bar q} & = & \frac{\alpha_s(\mu)}
{\pi} \int_{\tau_0}^1 d\tau
\sum_{q} \frac{d{\cal L}^{q\bar q}}{d\tau} \int_{\tau_0/\tau}^1 \frac{dz}{z}~
\hat \sigma_{\mathrm{LO}}(Q^2 = z \tau s)~\frac{32}{27} (1-z)^3.
\end{eqnarray}
The coefficient $C$ for the virtual corrections reads
\begin{eqnarray}
C & = & \pi^2 + c_1 + \frac{33-2N_F}{6} \log \frac{\mu^2}{Q^2} 
\nonumber \\
&& {} + \Re e~\frac{\int_{\hat t_-}^{\hat t_+} d\hat t \left\{c_2~C_\Box
(C_\triangle F_\triangle + C_\Box F_\Box) + c_3~\frac{p_T^2}{2 \hat t \hat u}
(Q^2-m_1^2-m_2^2) C^2_\Box G_\Box \right\}}
{\int_{\hat t_-}^{\hat t_+} d\hat t \left\{ |C_\triangle F_\triangle +
C_\Box F_\Box |^2 + |C_\Box G_\Box|^2 \right\}},
\label{eq:cfac}
\end{eqnarray}
where
\begin{equation}
\tau_0 = \frac{(m_1+m_2)^2}{s},  \qquad
p_T^2 = \frac{(\hat t-m_1^2)(\hat u-m_1^2)}{Q^2} - m_1^2. 
\end{equation}
The objects $P_{gg}(z), P_{gq}(z)$ denote the
Altarelli--Parisi splitting functions \cite{altpar}:
\begin{eqnarray}
P_{gg}(z) &=& 6\left\{ \left(\frac{1}{1-z}\right)_+
+\frac{1}{z}-2+z(1-z) \right\} + \frac{33-2N_F}{6}\delta(1-z), 
\nonumber \\
P_{gq}(z) &=& \frac{4}{3} \frac{1+(1-z)^2}{z},
\end{eqnarray}
where $N_F=5$ in our case.
The factorization scale of the parton--parton luminosities 
$d{\cal L}^{ij}/d\tau$ is denoted by $M$.
The coefficients $c_i$ for the individual final-state Higgs bosons
$\phi_1\phi_2$ are given by
\begin{equation}
\begin{array}[b]{lrclrclrcl}
\phi_1\phi_2=hh,hH,HH: \qquad & 
c_1 & = & \displaystyle \frac{11}{2}, \quad &
c_2 & = & \displaystyle \frac{4}{9}, \quad 
& c_3 & = & \displaystyle - \frac{4}{9}, \\ \\
\phi_1\phi_2=hA,HA: & 
c_1 & = & 6, & c_2 & = & \displaystyle \frac{2}{3}, & c_3 & = &
\displaystyle \frac{2}{3}~\frac{\hat t - \hat u}{Q^2 - m_1^2 - m_2^2}, \\ \\
\phi_1\phi_2=AA: & 
c_1 & = & \displaystyle \frac{11}{2}, & c_2 & = & -1, & c_3 & = & -1.
\end{array}
\end{equation}
In order to improve the validity of our results, we have inserted the full
expressions for the form factors $F_\triangle, F_\Box$ and $G_\Box$
in Eqs.~(\ref{eq:gghhlo}) and (\ref{eq:cfac}), 
i.e.\ including the exact dependence on the fermion masses.
This procedure is reasonable
since the QCD corrections are dominated by soft
and collinear gluon radiation, which do not resolve the Higgs couplings to
gluons, analogously to 
single-Higgs production via gluon fusion \cite{soft}.

A few remarks on the $s$-channel $Z$-boson exchange in $hA,HA$
production are in order. For the virtual corrections, the factorization
of the NLO corrections into the LO form factors and a universal
correction factor is exact for $m_t\to\infty$ and $m_b=0$. 
This is due to the fact that only the
pseudoscalar Goldstone component of the $Z$ boson contributes as in LO,
i.e.\ the QCD corrections coincide with the one to $s$-channel
pseudoscalar Higgs-boson exchange. For the real corrections, the
factorization is not exact, but the applied correction factor correctly
includes the dominant contributions, which are caused by soft and 
collinear gluon radiation. Additional infrared- and collinear-finite 
contributions, e.g.\ originating from $Zggg$ box corrections in
$gg\to g+Z^* \to g + hA,HA$ processes, are expected to be small, since they do
not exhibit large contributions from soft and collinear gluon radiation.
They are neglected in our analysis.
%

\subsection{Explicit expansion of the gluon-fusion diagrams}
 
We have derived the above results also by explicitly performing the 
heavy-mass expansion of the corresponding 
one- and two-loop Feynman 
diagrams. The amplitudes for the individual diagrams have been automatically
generated using the package {\sl FeynArts} \cite{FA}. The asymptotic expansion 
of the individual amplitudes in the heavy top-quark mass is carried out
directly in the integrand, i.e.\ before the integration over the
momentum space. We employ the general algorithm of Ref.~\cite{sm97} (see
also Ref.~\cite{asymp} and 
references therein) for the
asymptotic expansion of Feynman diagrams in dimensional regularization.
This method expresses the coefficients of the expansion in terms of simpler 
diagrams. At the one-loop level, this procedure leads to simple
one-loop vacuum integrals only. At the two-loop level, we get two-loop
vacuum integrals and products of one-loop vacuum integrals and massless
one-loop integrals with non-vanishing external momenta. The analytical
calculation of all those integrals is straightforward 
when using the Feynman-parameter technique. Since the
employed strategy leads to a very large number of terms in intermediate
steps, and since each step is algorithmic, we have fully automatized the
calculation in {\sl Mathematica} \cite{math}. In the following we sketch the
single steps of the calculation and give the results for the basic
integrals.

The general algorithm for the asymptotic expansion of any given Feynman 
graph $\Gamma$ in the limit $M_i\to\infty$ for some internal masses
$M_i$ can easily be summarized. Denoting the corresponding Feynman
amplitude by $F_\Gamma$ and the corresponding integrand by $I_\Gamma$,
the large-mass expansion reads
\def\asymp#1{\mathrel{\raisebox{-.4em}{$\widetilde{\scriptstyle #1}$}}}
\beq
F_\Gamma \; = \; \int \biggl(\prod_l d^n q_l\biggr) \, I_\Gamma
\;\; \asymp{M_i\to\infty} \;\; 
\sum_\gamma \, \int \biggl(\prod_l d^n q_l\biggr) \,  
I_{\Gamma/\gamma} \, {\cal T}_{p^\gamma_i,m_i} I_\gamma,
\label{eq:Fexpand}
\eeq
where $q_l$ are the integration momenta.
The sum on the r.h.s.\ runs over all subgraphs $\gamma$ of $\Gamma$ which 
contain all propagators with the heavy masses $M_i$ and which are 
irreducible with respect to those lines of $\gamma$ 
that carry light masses 
$m_i$. The integrand of the subgraph $\gamma$ is denoted by $I_\gamma$.
The reduced graph $\Gamma/\gamma$ results from $\Gamma$ upon
shrinking $\gamma$ to a point, and the integrand $I_{\Gamma/\gamma}$ is
defined such that $I_\Gamma=I_\gamma I_{\Gamma/\gamma}$. The symbol
${\cal T}_{p^\gamma_i,m_i}$ represents an operator that replaces the
integrand $I_\gamma$ by its Taylor series in the expansion parameters
$p^\gamma_i$ and $m_i$, where $p^\gamma_i$ are the external momenta of
the subgraph $\gamma$. Therefore, Eq.~(\ref{eq:Fexpand}) expresses the
original integral $F_\Gamma$ in terms of an infinite sum over simpler
integrals. For any given power $\xi^a$, this sum contains only a finite
number of terms that are non-vanishing in 
$(F_\Gamma|_{M_i\to\xi M_i})/\xi^a$ after the scaling limit $\xi\to\infty$ 
is taken. These terms can 
easily be determined by power counting.
This general strategy for the heavy-mass expansion will become more
transparent when 
we inspect in more detail the different types of graphs 
that are relevant in our case. 

We start by considering the relevant one-loop integrals in the limit
$m_t\to\infty$. They contain only top-quark propagators in the loop, both 
for the LO calculation 
and for the real NLO corrections. According 
to the algorithm (\ref{eq:Fexpand}), there is only one relevant subgraph 
$\gamma$, namely the subgraph $\gamma_{\mathrm{loop}}$ containing only the 
propagators of lines inside the loop.
If $\Gamma$ is irreducible, we have $\Gamma=\gamma_{\mathrm{loop}}$. 
The Taylor-expansion operator ${\cal T}$ replaces each propagator 
$P(q-p,m_t)$ by 
\begin{equation}
P(q-p,m_t) = [(q-p)^2-m_t^2]^{-1} =
\sum_{l=0}^\infty (q^2-m_t^2)^{-1-l} (2q p-p^2)^l,
\end{equation}
where $q$ denotes the integration momentum, and $p$ is any combination
of external momenta. These replacements express each one-loop diagram
by a sum of terms containing one-loop vacuum integrals
\begin{equation}
V^{(1)}_{\mu_1\dots\mu_R}(n_1;m_1) = 
\frac{(2\pi\mu)^{4-n}}{i\pi^2} \int d^n q \,
\frac{q_{\mu_1}\dots q_{\mu_R}}{(q^2-m_1^2)^{n_1}}. 
\label{eq:V1tens}
\end{equation}
The terms that are non-vanishing in the heavy-mass limit can be
determined by simple power counting, since an explicit factor of $m_t$ and 
the integration momentum $q$ contribute to the scaling factor in $m_t$
exactly in the same way. All non-vanishing vacuum tensor integrals 
(\ref{eq:V1tens}) can be decomposed into terms that are products of 
metric tensors $g_{\mu\nu}$ and coefficient factors. The coefficients 
for the different covariants, which span the whole tensor, can be 
algebraically expressed in terms of scalar vacuum integrals. This algebraic 
reduction, which proceeds recursively in the tensor rank, is standard. 
The trick is to contract the equation that expresses the integral in
terms of covariants with a set of some 
suitably chosen covariants.
On the side of the integral, this leads to integrals that are already
known; on the other side of this equation, one gets linear combinations 
for the tensor coefficients. The coefficients are obtained by inverting 
a system of such linear equations for the coefficients. 
For the case of one-loop tensor vacuum integrals, 
at most a single covariant structure contributes,
namely the totally symmetric tensor built of metric tensors $g_{\mu\nu}$.
Only scalar one-loop integrals, i.e.\ the ones of (\ref{eq:V1tens}) with 
$R=0$, have to be computed explicitly. A simple calculation yields
\begin{equation}
V^{(1)}(n_1;m_1) = 
(-1)^{n_1} (4\pi\mu^2)^{\frac{4-n}{2}} m^{n-2n_1} 
\frac{ \Gamma\left(n_1-\frac{n}{2}\right) } { \Gamma(n_1) }.
\end{equation}

\begin{figure}[t]
\begin{picture}(180,110)(-20,-20)
\SetScale{0.8}
\Gluon(0,100)(50,100){3}{6}
\Gluon(0,0)(50,0){3}{6}
\Gluon(95,0)(95,100){3}{12}
\ArrowLine(50,100)( 95,100)
\ArrowLine(95,100)(140,100)
\ArrowLine(140,100)(140,0)
\ArrowLine(140,0)(95,0)
\ArrowLine( 95,0)(50,0)
\ArrowLine(50,0)(50,100)
\DashLine(140,100)(190,100){5}
\DashLine(140,0)(190,0){5}
\put(165,76){$\phi_1$}
\put(165,-5){$\phi_2$}
\put(-15,-3){$g$}
\put(-15,78){$g$}
\put(120,38){$t$}
\put(10,-20){(a)}
\SetScale{1}
\end{picture}
\hspace*{3em}
\begin{picture}(180,110)(-20,-20)
\SetScale{0.8}
\Gluon(0,100)(50,100){3}{6}
\Gluon(0,0)(50,0){3}{6}
\Gluon(50,100)( 95,100){3}{6}
\Gluon(50,50)(50,100){3}{6}
\ArrowLine(50,0)(50,50)
\ArrowLine(50,50)(95,100)
\ArrowLine(95,100)(140,100)
\ArrowLine(140,100)(140,0)
\ArrowLine(140,0)(50,0)
\DashLine(140,100)(190,100){5}
\DashLine(140,0)(190,0){5}
\put(165,76){$\phi_1$}
\put(165,-5){$\phi_2$}
\put(-15,-3){$g$}
\put(-15,78){$g$}
\put(120,38){$t$}
\put(10,-20){(b)}
\SetScale{1}
\end{picture}
\\[2em]
\begin{picture}(180,110)(-20,-20)
\SetScale{0.8}
\Gluon(0,100)(50,50){3}{8}
\Gluon(0,0)(50,50){3}{8}
\Gluon(50,50)( 95,100){3}{8}
\Gluon(50,50)( 95,0){3}{8}
\ArrowLine(95,0)(95,100)
\ArrowLine(95,100)(140,100)
\ArrowLine(140,100)(140,0)
\ArrowLine(140,0)(95,0)
\DashLine(140,100)(190,100){5}
\DashLine(140,0)(190,0){5}
\put(165,76){$\phi_1$}
\put(165,-5){$\phi_2$}
\put(-15,-3){$g$}
\put(-15,78){$g$}
\put(120,38){$t$}
\put(10,-20){(c)}
\SetScale{1}
\end{picture}
\hspace*{3em}
\begin{picture}(180,110)(-20,-20)
\SetScale{0.8}
\Gluon(0,100)(50,100){3}{6}
\Gluon(0,0)(50,0){3}{6}
\ArrowLine(95,0)(95,100)
\Gluon(50,100)( 95,100){3}{6}
\ArrowLine(95,100)(140,100)
\ArrowLine(140,100)(140,0)
\ArrowLine(140,0)(95,0)
\Gluon(50,0)( 95,0){3}{6}
\Gluon(50,0)(50,100){3}{12}
\DashLine(140,100)(190,100){5}
\DashLine(140,0)(190,0){5}
\put(165,76){$\phi_1$}
\put(165,-5){$\phi_2$}
\put(-15,-3){$g$}
\put(-15,78){$g$}
\put(120,38){$t$}
\put(10,-20){(d)}
\SetScale{1}
\end{picture}
\caption[]{\it \label{fg:twoloopdia} Typical two-loop box diagrams with
one, two, or three internal gluons.}
\end{figure}
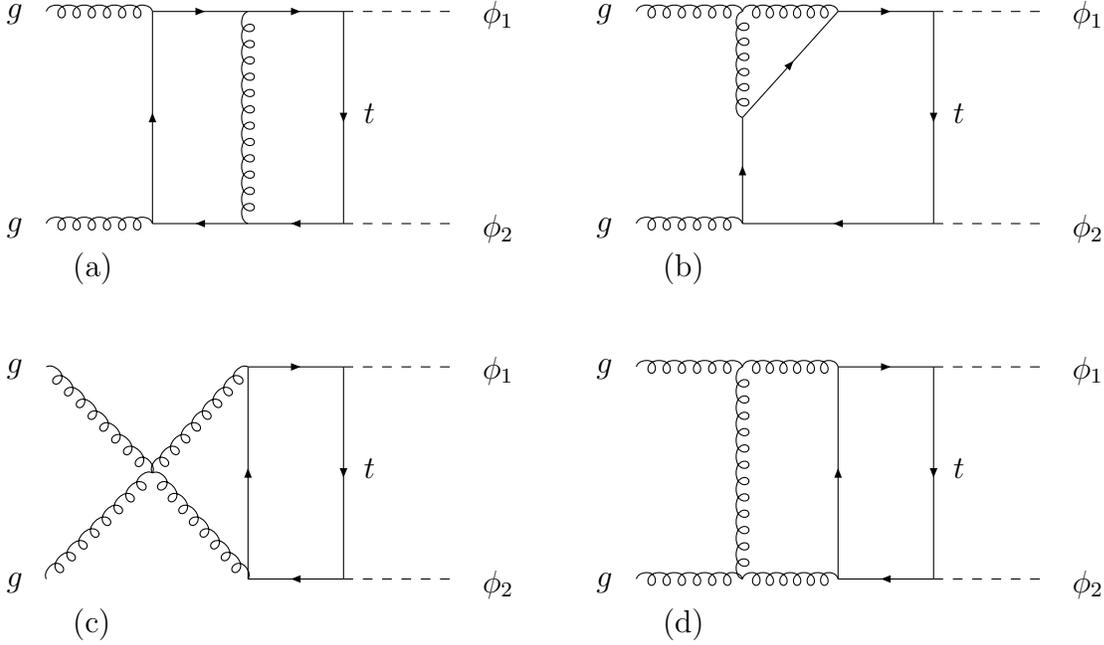

At the two-loop level, there are two basically different types of diagrams.
The first type contains two independent top-quark loops. Such
diagrams do not lead to genuine two-loop integrals and can be treated like 
the one-loop diagrams above. Topologically those diagrams are represented 
by the third graph of Fig.~\ref{fg:gghhdiaeff}. The second type of graphs 
is formed by the genuine two-loop diagrams. Each of those diagrams contains 
a closed top-quark loop and one, two, or three internal gluon lines. 
Typical box diagrams are shown in Fig.~\ref{fg:twoloopdia}.
For all such genuine two-loop diagrams there are two subgraphs $\gamma$
that are relevant in the expansion (\ref{eq:Fexpand}) for $m_t\to\infty$.
The first subgraph is the diagram $\gamma_{\mathrm{loop}}$ built of all
lines inside the loops, the second is 
given by the closed top-quark loop.

First we consider the case $\gamma=\gamma_{\mathrm{loop}}$. The Taylor 
expansion of the integrand $I_{\gamma_{\mathrm{loop}}}$ involves the 
consistent expansion of each propagator about the external momenta of
the process. This means that each propagator $P(q-p,m)$ is replaced by
\begin{equation}
P(q-p,m) = [(q-p)^2-m^2]^{-1} =
\sum_{l=0}^\infty (q^2-m^2)^{-1-l} (2q p-p^2)^l, \qquad m=0,m_t,
\end{equation}
where $q$ is a linear combination of the two integration momenta $q_1$ 
and $q_2$, and $p$ consists of external momenta $p_i$. This replacement 
leads to two-loop vacuum integrals of the general form 
\begin{eqnarray}
\lefteqn{V^{(2)}_{\mu_1\dots\mu_R;\nu_1\dots\nu_S}(n_1,n_2,n_3;m_1,m_2,m_3)}
\nonumber\\[.3em]
&=& 
\frac{(2\pi\mu)^{4-n}}{i\pi^2} \int d^n q_1 \,
\frac{(2\pi\mu)^{4-n}}{i\pi^2} \int d^n q_2 \,
\frac{q_{1,\mu_1}\dots q_{1,\mu_R} \, q_{2,\nu_1}\dots q_{2,\nu_S}}
{(q_1^2-m_1^2)^{n_1} (q_2^2-m_2^2)^{n_2} [(q_1+q_2)^2-m_3^2]^{n_3}}, 
\hspace*{2em}
\end{eqnarray}
with the actual mass insertions $m_1=0$ and $m_2=m_3=m_t$.
In the power counting, which determines the non-vanishing terms for
$m_t\to\infty$, factors of the integration momenta
$q_1$ and $q_2$ contribute in the same way as explicit factors of $m_t$.
The algebraic reduction of two-loop tensor vacuum integrals proceeds
along the same lines as described for the one-loop case above. The only
difference is that tensors need not be totally symmetric beyond one
loop. Finally, we are left with 
the scalar integral $(R=S=0)$,
which can be easily calculated, 
\begin{eqnarray}
V^{(2)}(n_1,n_2,n_3;0,m,m) &=& 
(-1)^{n_1+n_2+n_3} (4\pi\mu^2)^{4-n} (m^2)^{n-n_1-n_2-n_3} 
\nonumber\\
&& \hspace*{-6em} {} \times
\frac{ \Gamma(n_1+n_2+n_3-n) \Gamma\left(\frac{n}{2}-n_1\right) 
\Gamma\left(n_1+n_2-\frac{n}{2}\right) \Gamma\left(n_1+n_3-\frac{n}{2}\right) }
{ \Gamma\left(\frac{n}{2}\right) \Gamma(n_2) \Gamma(n_3)
\Gamma(2n_1+n_2+n_3-n) }.
\hspace*{2em}
\end{eqnarray}

Now we identify the subgraph $\gamma$ with the top-quark loop. In this
case the Taylor expansion in the integrand $I_\gamma$ concerns the momenta 
that are external with respect to the top-quark loop. Thus, all
top-quark propagators $P(q_2-p,m_t)$ are replaced by
\begin{equation}
P(q_2-p,m_t) = [(q_2-p)^2-m_t^2]^{-1} =
\sum_{l=0}^\infty (q_2^2-m_t^2)^{-1-l} (2q_2 p-p^2)^l,
\end{equation}
where $q_2$ is the loop momentum running through the top-quark loop.
Note that $p$ includes all external momenta of the process as well as 
the loop momentum $q_1$ running through the internal gluon lines. The
integration over $q_2$ leads to one-loop vacuum integrals of the form
(\ref{eq:V1tens}), the calculation of which is described above.
The integration over $q_1$ involves only massless propagators to the
first power. Since the $q_1$ integration does not involve any
$m_t$-terms, it does not affect the power counting in $m_t$ at all.
For one-gluon exchange the integral over $q_1$ is a massless 
tadpole, which vanishes in dimensional regularization. For two-gluon and
three-gluon exchange the $q_1$ integration leads to the one-loop tensor
integrals
\begin{eqnarray}
B_{\mu_1\dots\mu_R}(p) &=& 
\frac{(2\pi\mu)^{4-n}}{i\pi^2} \int d^n q_1 \,
\frac{q_{1,\mu_1}\dots q_{1,\mu_R}} {q_1^2(q_1+p)^2}, 
\nonumber\\[.5em]
C_{\mu_1\dots\mu_R}(p_1,p_2) &=& 
\frac{(2\pi\mu)^{4-n}}{i\pi^2} \int d^n q_1 \,
\frac{q_{1,\mu_1}\dots q_{1,\mu_R}} {q_1^2(q_1+p_1)^2(q_1+p_2)^2}. 
\end{eqnarray}
The tensor integrals can again be recursively reduced to the
corresponding scalar integrals in a fully algebraic manner.
The relevant scalar integrals can be easily calculated and are given by
\begin{eqnarray}
B(p)\Big|_{p^2\ne 0} &=& 
\left(\frac{4\pi\mu^2}{-p^2-i0}\right)^{\frac{4-n}{2}} \,
\frac{ \Gamma\left(2-\frac{n}{2}\right) \Gamma\left(\frac{n}{2}-1\right)^2 }
{ \Gamma(n-2) },
\nonumber\\[.5em]
C(p_1,p_2)\Big|_{p_1^2=p_2^2=0,\, p^2=(p_1-p_2)^2\ne 0} &=&
\frac{1}{p^2} \, \left(\frac{4\pi\mu^2}{-p^2-i0}\right)^{\frac{4-n}{2}} \,
\frac{ \Gamma\left(3-\frac{n}{2}\right) \Gamma\left(\frac{n}{2}-2\right)^2 }
{ \Gamma(n-3) }.
\end{eqnarray}
For $p^2=0$ the integrals are zero in dimensional regularization. The
case $B(p)$ with $p^2=0$ occurs, for instance, 
in graphs like Fig.~\ref{fg:twoloopdia}b. 
Therefore, we find that the contribution 
in the expansion (\ref{eq:Fexpand}) for which $\gamma$ is the top-quark loop
is only non-vanishing in diagrams like Figs.~\ref{fg:twoloopdia}c and
d, where both external gluons are attached to the internal gluon
lines. 

A few more ``physical'' remarks on the formally described algorithm for
the asymptotic expansion seem to be in order. The different
contributions to a given Feynman graph that are associated 
with the subgraphs $\gamma$ in the expansion (\ref{eq:Fexpand}) are
directly related to the effective diagrams in the approach 
of the low-energy theorem described above. Shrinking the subgraph $\gamma$ 
to a point leads to the corresponding effective diagram, where the point
arising from $\gamma$ is the point-like interaction of the effective 
Lagrangian. A non-vanishing contribution of a subgraph 
$\gamma\ne\gamma_{\mathrm{loop}}$ requires that at least one external 
momentum $p_i$ passes through a massless propagator; otherwise the
loop integral involving the massless propagators is zero. In other words, 
it is necessary that there exists
a cut through the diagram that passes
exclusively massless lines. Therefore, only diagrams with such a 
``massless cut'' can lead to contributions to effective diagrams in which 
an effective coupling appears in loops. But all diagrams in general 
contribute to tree-like effective diagrams, which result from shrinking
the complete loop part $\gamma_{\mathrm{loop}}$ to a point.

Finally, we mention that $\gamma_5$, which appears in the case of
pseudoscalar Higgs bosons, is treated according to the prescription of
't~Hooft and Veltman \cite{dimreg}. Technically we substitute $\gamma_5$
by $i\epsilon^{\mu_1\mu_2\mu_3\mu_4}\gamma_{\mu_1}\gamma_{\mu_2}
\gamma_{\mu_3}\gamma_{\mu_4}/4!$ before the evaluation of the Dirac
trace so that the actual trace calculation can be carried out for usual
$n$-dimensional Dirac matrices. The correct projection of the trace result
on the physical four-dimensional space is achieved upon the contraction
with the four-dimensional 
$\epsilon$-tensor.
In this approach, all loop integrations can be carried out before the
contraction with $\epsilon$, i.e.\ $n$-dimensional momenta can be used.
Note, however, that the contraction with $\epsilon$ necessarily occurs
before the integration over the phase space of the radiated parton in
the real corrections, i.e.\ one has to take care of the
four-dimensionality of $\epsilon$ there.
Moreover, an additional spurious counter term has to be added to the $At\bar t$
vertex 
(see also \cite{higgsqcd,agg}).

\subsection{Drell--Yan-like processes}
\begin{figure}[hbt]
\begin{center}
\begin{picture}(160,100)(0,0)

\ArrowLine(0,80)(50,50)
\ArrowLine(50,50)(0,20)
\Photon(50,50)(100,50){3}{5}
\DashLine(100,50)(150,80){5}
\DashLine(100,50)(150,20){5}
\put(155,16){$h,H$}
\put(-15,18){$\bar q$}
\put(-15,78){$q$}
\put(70,60){$Z$}
\put(155,76){$A$}

\end{picture}  \\
\setlength{\unitlength}{1pt}
\caption[ ]{\label{fg:dy} \it Diagram contributing to $q\bar q \to Ah,AH$
at lowest order.}
\end{center}
\end{figure}
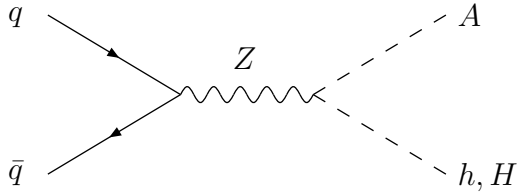
Pairs of scalar and pseudoscalar Higgs bosons can also be produced in $q\bar q$
collisons via $s$-channel $Z$-boson exchange, see
Fig.\ \ref{fg:dy}. At LO the partonic cross sections are given by
\begin{equation}
\hat\sigma_{\mathrm{LO}}(q\bar q\to Ah,AH)=\lambda^2_{ZAh,ZAH}
\frac{G_F^2 M_Z^4}{288\pi}(v_q^2 + a_q^2)
\frac{\lambda(Q^2,M_A^2,M_{h,H}^2)^\frac{3}{2}}{(Q^2)^2(Q^2-M_Z^2)^2} \, ,
\end{equation}
where $Q^2$ denotes the partonic c.m.\ energy squared, and $v_q=2I_{3q}-4e_q
\sin^2\theta_W$, $a_q = 2I_{3q}$ are the vectorial and axial charges of the
initial-state quarks. The QCD corrections
coincide analytically with the QCD corrections to
the Drell--Yan process $q\bar q \to Z^*$, if squarks and gluinos are heavy, so
that their contributions can be neglected. Thus the NLO cross section can be
expressed as:
\begin{eqnarray}
\sigma_{\mathrm{NLO}}(pp\to Ah,AH+X) & = & 
\sigma_{\mathrm{LO}} + \Delta\sigma_{q\bar q} +
\Delta\sigma_{qg}, \nonumber \\
\sigma_{\mathrm{LO}} & = & \int_{\tau_0}^1 d\tau \sum_q
\frac{d{\cal L}^{q\bar q}}{d\tau} 
\hat\sigma_{\mathrm{LO}}(Q^2 = \tau s), \nonumber \\
\Delta\sigma_{q\bar q} & = & \frac{\alpha_s(\mu)}{\pi} \int_{\tau_0}^1 d\tau
\sum_q \frac{d{\cal L}^{q\bar q}}{d\tau} \int_{\tau_0/\tau}^1 dz~\hat
\sigma_{\mathrm{LO}}(Q^2 = \tau z s)~\omega_{q\bar q}(z), \nonumber \\
\Delta\sigma_{qg} & = & \frac{\alpha_s(\mu)}{\pi} \int_{\tau_0}^1 d\tau
\sum_{q,\bar q} \frac{d{\cal L}^{qg}}{d\tau} \int_{\tau_0/\tau}^1 dz~\hat
\sigma_{\mathrm{LO}}(Q^2 = \tau z s)~\omega_{qg}(z),
\hspace*{2em}
\eea
with the coefficient functions \cite{dyqcd}
\bea
\omega_{q\bar q}(z) & = & -P_{qq}(z) \log \frac{M^2}{\tau s}
+ \frac{4}{3}\left\{ 2[\zeta_2-2]\delta(1-z) + 4\left(\frac{\log (1-z)}{1-z}
\right)_+
\phantom{\frac{M^2}{s}}\!\!\!\!\!\!\!\!\!\!
\right. \nonumber \\
& & \hspace*{7cm} \left. {}- 2(1+z)\log(1-z)
\phantom{\frac{M^2}{s}}\!\!\!\!\!\!\!\!\!\!
\right\}, \nonumber \\
\omega_{qg}(z) & = & -\frac{1}{2} P_{qg}(z) \log \left(
\frac{M^2}{(1-z)^2 \tau s} \right) + \frac{1}{8}\left\{ 1+6z-7z^2 \right\}.
\end{eqnarray}
The Altarelli--Parisi splitting functions $P_{q\bar q}$ and $P_{qg}$ are given
by \cite{altpar}
\begin{eqnarray}
P_{qq}(z) & = & \frac{4}{3} \left\{ 2\left(\frac{1}{1-z}\right)_+-1-z
+\frac{3}{2}\delta(1-z) \right\}, \nonumber \\
P_{qg}(z) & = & \frac{1}{2} \left\{ z^2 + (1-z)^2 \right\}.
\end{eqnarray}
 
\section{Results}
\label{se:results}
\subsection{Standard Model}
The analysis for SM 
Higgs-boson pair production has been carried out for the
LHC c.m.\ energy $\sqrt{s}=14$~TeV. 
The top-quark and bottom-quark masses have been chosen as $m_t=175$~GeV and
$m_b=5$~GeV. 
We have adopted the CTEQ4L and CTEQ4M \cite{cteq4} parton densities for
the LO and NLO cross sections, respectively, corresponding to the QCD
parameters $\Lambda_5^{\mathrm{LO}}=181$~MeV and $\Lambda_5^{\msb}=202$~MeV.
Since our results have been obtained in the heavy-quark limit, they
can be expected to be reliable for $M_H\lsim 200$~GeV, based on the experience
from single-Higgs production via gluon fusion.

\begin{figure}[p]
\vspace*{0.5cm}
\hspace*{0.5cm}
\begin{turn}{-90}%
\epsfxsize=9cm \epsfbox{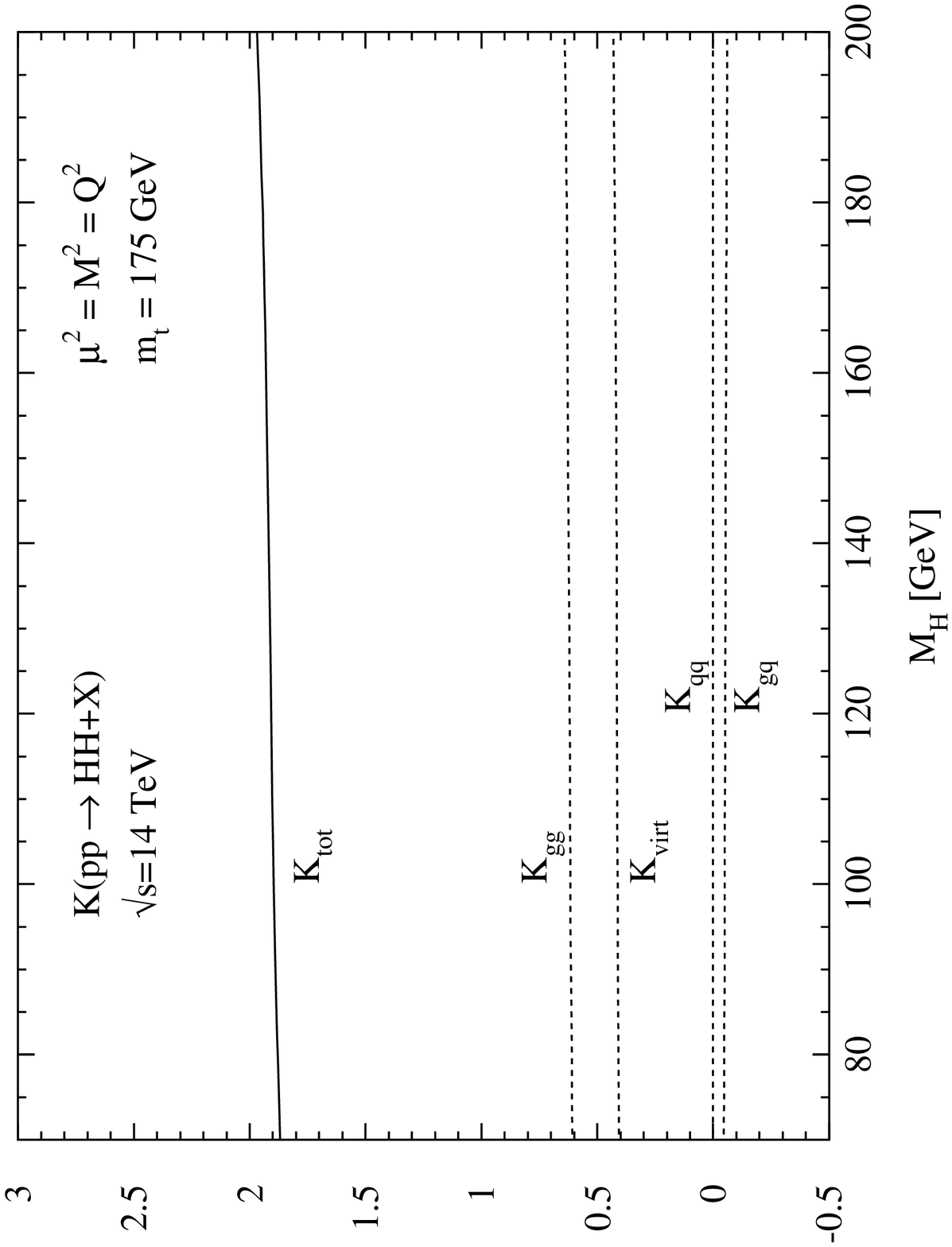}
\end{turn}
\caption[]{\it \label{fg:ksm} K factors of the QCD-corrected gluon-fusion
SM cross section $\sigma(pp \to HH+X)$ at the LHC with c.m.~energy $\sqrt{s}=14$
TeV. The dashed lines show the individual contributions of the four terms of
the QCD corrections given in Eq.~(\ref{eq:gghqcd}). 
}
\vspace*{0.3cm}
\hspace*{0.5cm}
\begin{turn}{-90}%
\epsfxsize=9cm \epsfbox{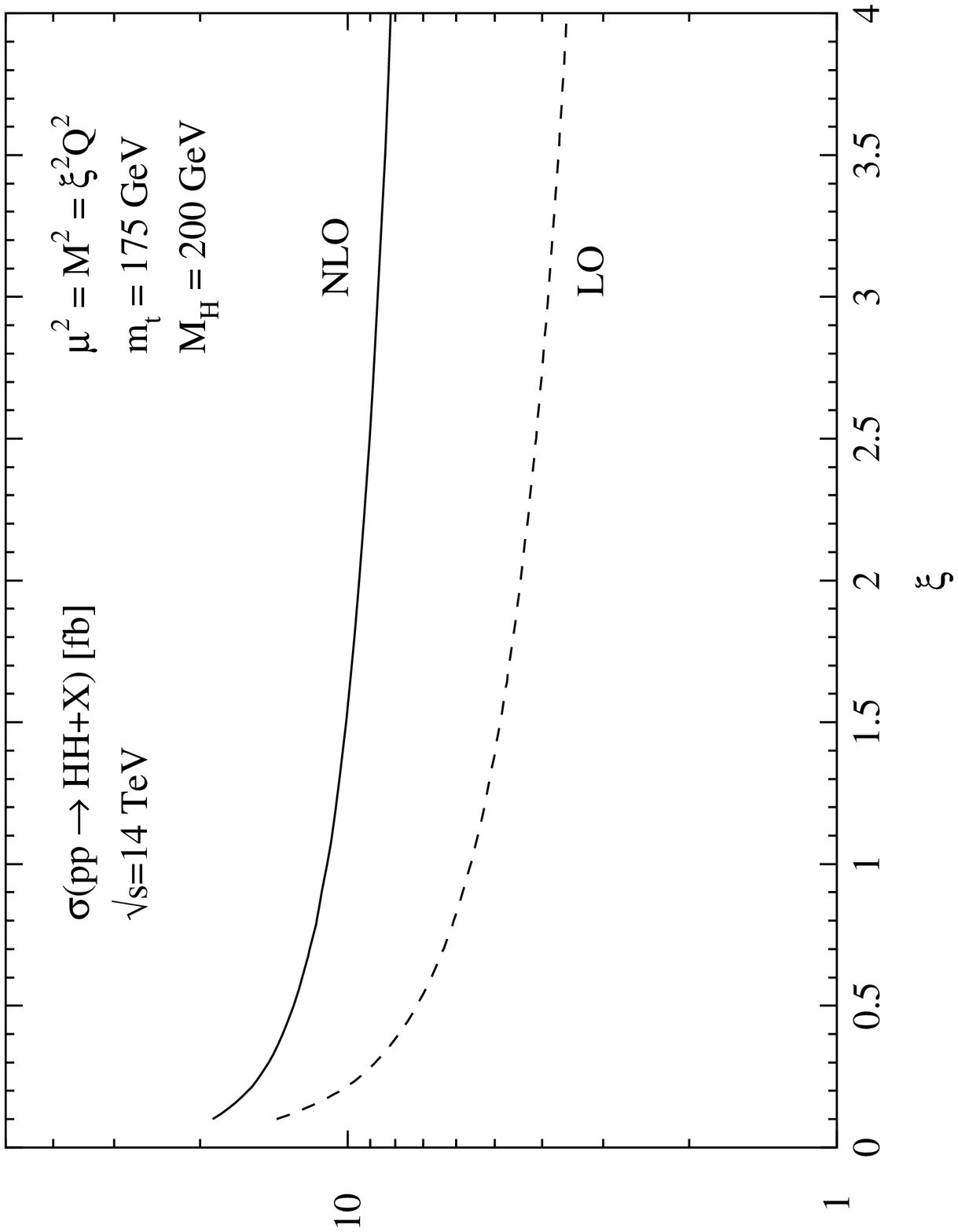}
\end{turn}
\caption[]{\it \label{fg:scalesm} The renormalization and factorization scale
dependence of the SM-Higgs pair-production cross section at LO and
NLO for a Higgs mass $M_H = 200$ GeV.}
\end{figure}
In order to investigate the size of the QCD corrections,
we define the $K$
factor $K=\sigma_{\mathrm{NLO}}/\sigma_{\mathrm{LO}}$ as the ratio of the NLO and LO cross
sections, where the parton densities and the strong coupling $\alpha_s$ are
taken at NLO and LO, respectively.  The $K$ factor for \
SM-Higgs pair production is presented
in Fig.~\ref{fg:ksm} as a function of the Higgs mass $M_H$
and shows little variation with $M_H$.
It ranges between about 1.9 and 
2.0, thus enhancing the LO cross section significantly. Moreover,
the broken lines of Fig.~\ref{fg:ksm} show the contributions of the individual
terms of Eq.~(\ref{eq:gghqcd}). It can be clearly inferred that analogously
to the case of single-Higgs production the 
(infrared-regularized) virtual
and real corrections originating from $gg$ intial states dominate the
QCD corrections, while the $gq$ and $q\bar q$ initial states do not exceed
about 5\% in total. We note that the 
values shown for $K_{\mathrm{virt}}$,
$K_{gg}$, $K_{gq}$, and $K_{qq}$ do not
exactly add up to $K_{\mathrm{tot}}-1$, since the individual
contributions are obtained by taking 
NLO parton densities and the strong coupling in the NLO cross sections, but
LO quantities in the LO cross sections consistently.

In order to
investigate the reliability of our results, the residual dependence of the
cross section on the renormalization and factorization scales is shown in
Fig.~\ref{fg:scalesm} for a Higgs mass of $M_H=200$ GeV at LO and NLO.
The scale dependence 
significantly decreases 
compared with the LO result by including the QCD corrections. 
However, the still monotonic decrease of the NLO cross section with 
increasing scales 
signals the need for further improvements.
Nevertheless, the theoretical uncertainty of the NLO result can be estimated
to about 20\% from the residual scale dependence.

\begin{figure}[p]
\hspace*{0.5cm}
\begin{turn}{-90}%
\epsfxsize=9cm \epsfbox{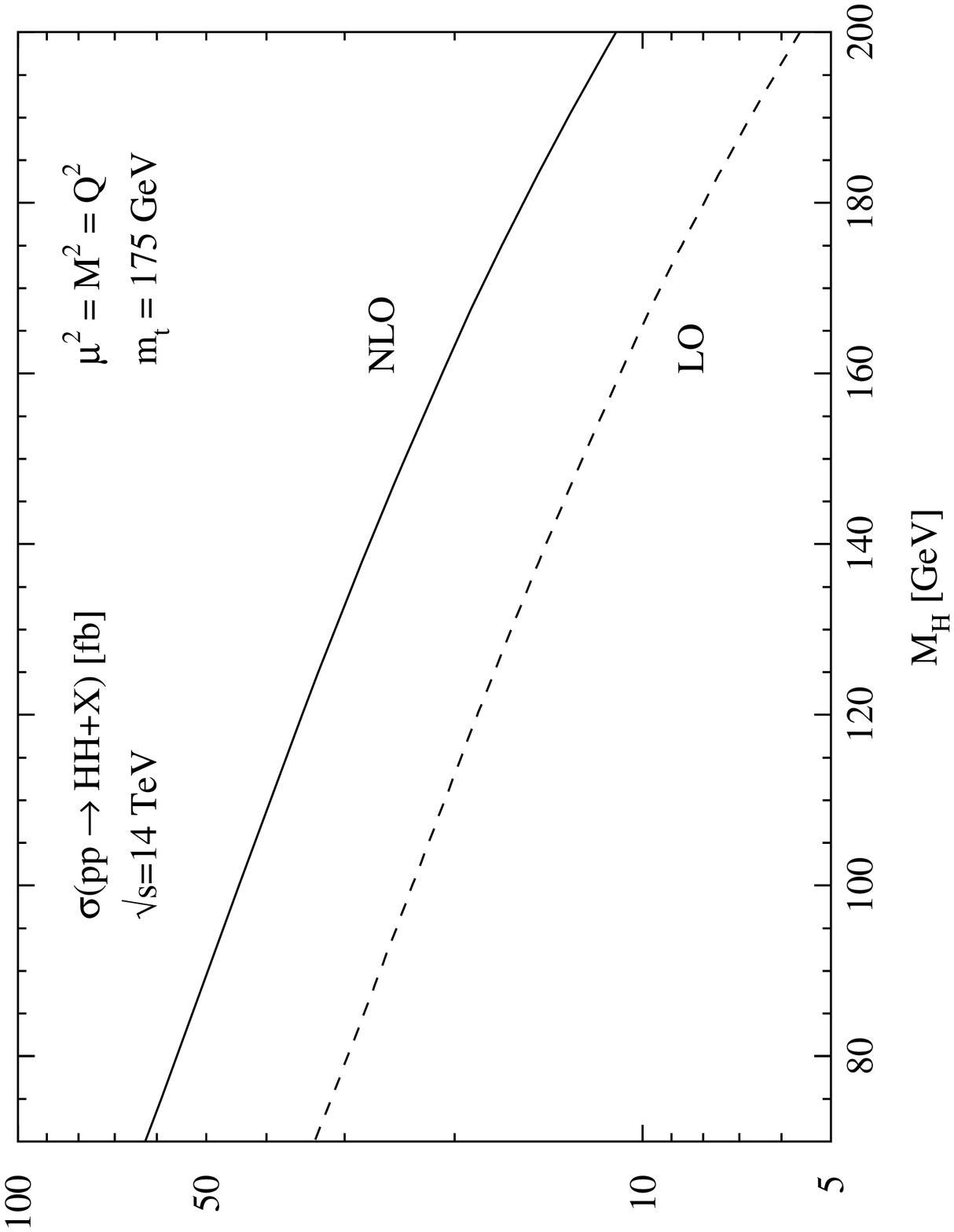}
\end{turn}
\caption[]{\it \label{fg:cxnsm} SM prediction of the Higgs-boson 
pair-production 
cross section at LO and NLO as a function of the Higgs mass $M_H$.}
\vspace*{0.5cm}
\hspace*{1.5cm}
\begin{turn}{-90}%
\epsfxsize=9cm \epsfbox{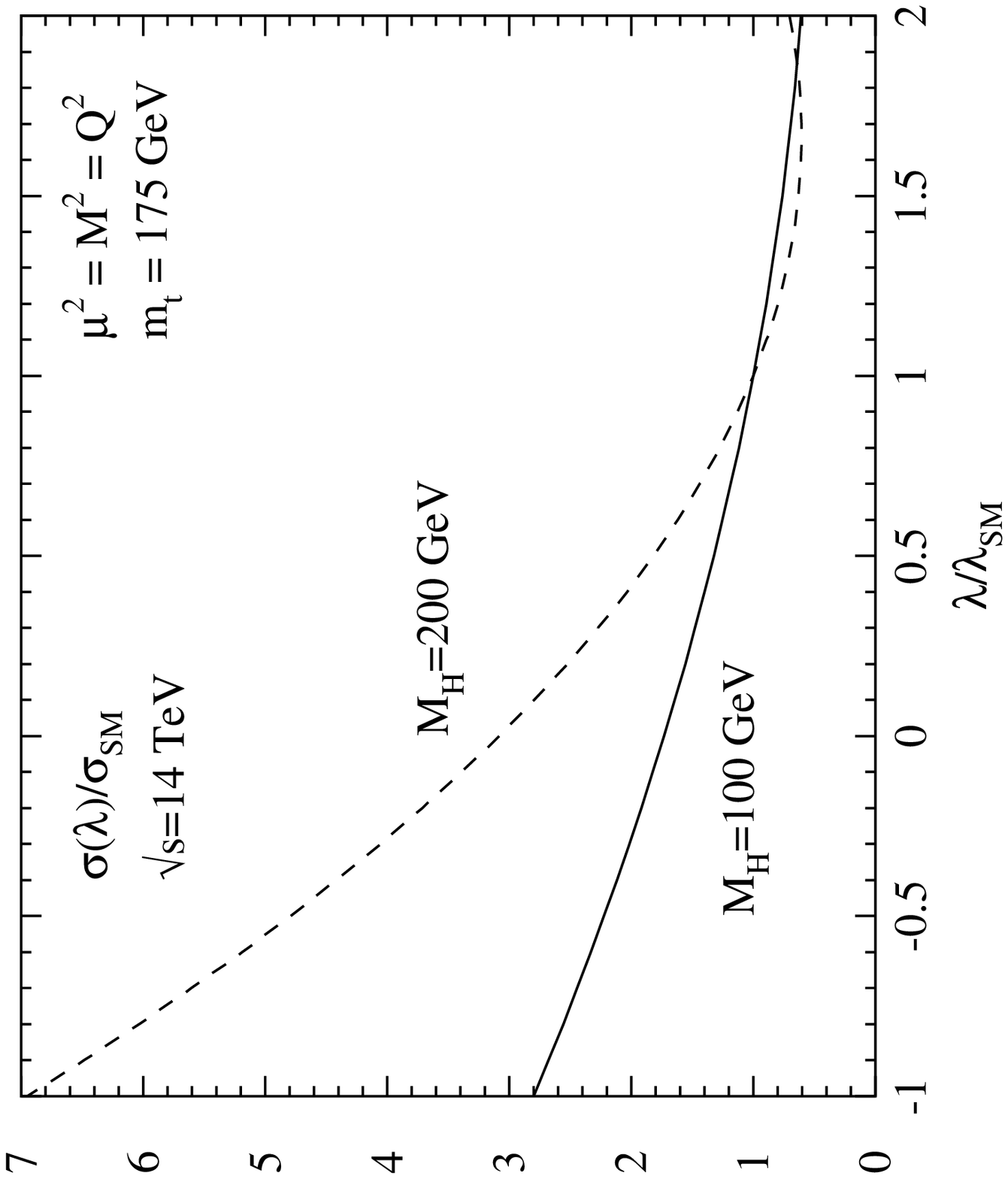}
\end{turn}
\caption[]{\it \label{fg:tria} Ratio of the Higgs-boson pair-production cross
section at NLO with a non-standard trilinear coupling $\lambda$ and the 
SM cross section as a function of the trilinear coupling in units of the SM 
one.}
\end{figure}

The final results for the total Higgs 
pair-production cross sections at LO and NLO are presented in
Fig.~\ref{fg:cxnsm} as a function of the Higgs mass $M_H$.
We recall that the full mass dependence of the LO form factors is included
in all numerical evaluations, i.e.\ only the relative NLO correction is
treated in the heavy-quark limit,
in order to increase the validity of our results.  
For $M_H\lsim 200$~GeV the cross section exceeds 10~fb, 
leading to more than about 3000 events at the LHC, once the anticipated
integrated luminosity 
$\int {\cal L} dt = 3 \times 10^5~$pb$^{-1}$ is reached%
\footnote{The
cross section at the Tevatron is less than 0.2~fb for $\sqrt{s}=2$~TeV.}.
The typical signatures for Higgs-boson pair production are $b\bar b b\bar b$ 
and $b\bar b\tau^+ \tau^-$ final states for $M_H\lsim 140$~GeV.
For $M_H\gsim 160$~GeV, the Higgs-boson pairs decay predominantly
into four vector bosons.

Finally, we present the sensitivity of the SM-Higgs pair-production cross
section to the trilinear Higgs coupling in Fig.~\ref{fg:tria}, which shows
the ratio of the total cross section with a non-standard trilinear coupling
$\lambda$ and the SM one as a function of the trilinear coupling,
varying in units of the SM coupling. The cross section becomes
significantly larger for smaller trilinear couplings,
so that this process may
serve as a possibility to measure this coupling and test the SM Higgs sector,
if the signal can be extracted from the QCD background.

The trilinear Higgs coupling can potentially be measured at high-energy
$e^+e^-$ colliders \cite{eehhsm}. Since the rate for multiple Higgs-boson 
production at $e^+e^-$ colliders is small, high luminosities are required
for a sufficient number of events in order to probe this coupling. However, the
backgrounds will be well under control.

\subsection{Minimal supersymmetric extension}
Promising MSSM-Higgs 
pair-production processes at the LHC are $hh$, $hH$,
$HH$, $hA$ and $HA$ production with sizeable ranges in
the parameter space where the cross sections exceed 10~fb 
\cite{gghhlo}. The $gg\to hh$ process can be used to cover a part of the
MSSM parameter space for small $\tgb$ via its decay modes into
$b\bar b\gamma\gamma$ final states \cite{richter} in the region 
where this process is dominated by resonant $gg\to H\to hh$ production.
Since our calculation is expected to be valid for small values of
$\tgb$ (where the 
$b$-quark contribution can be neglected)
and Higgs masses below the $t\bar t$ threshold, it may be assumed to
reliably approximate the cross sections of light-scalar-Higgs pair-production
in particular. Generally, in the MSSM, 
we expect our results to be valid for
$M_A \lsim 200$ GeV, while for larger pseudoscalar masses 
at least the heavy $s$-channel Higgs particles become too heavy with 
respect to the top-quark mass.

In our analysis we have included the MSSM Higgs masses and couplings at the
two-loop level in the effective-potential 
approach \cite{mssmrad2}. Moreover,
we have included all available higher-order corrections to the total MSSM Higgs
decay widths 
\cite{higgsrev}, which appear in the $s$-channel Higgs propagators of the
triangular loop contributions.

\begin{figure}[t]
\vspace*{0.0cm}
\hspace*{0.0cm}
\begin{turn}{-90}%
\epsfxsize=10cm \epsfbox{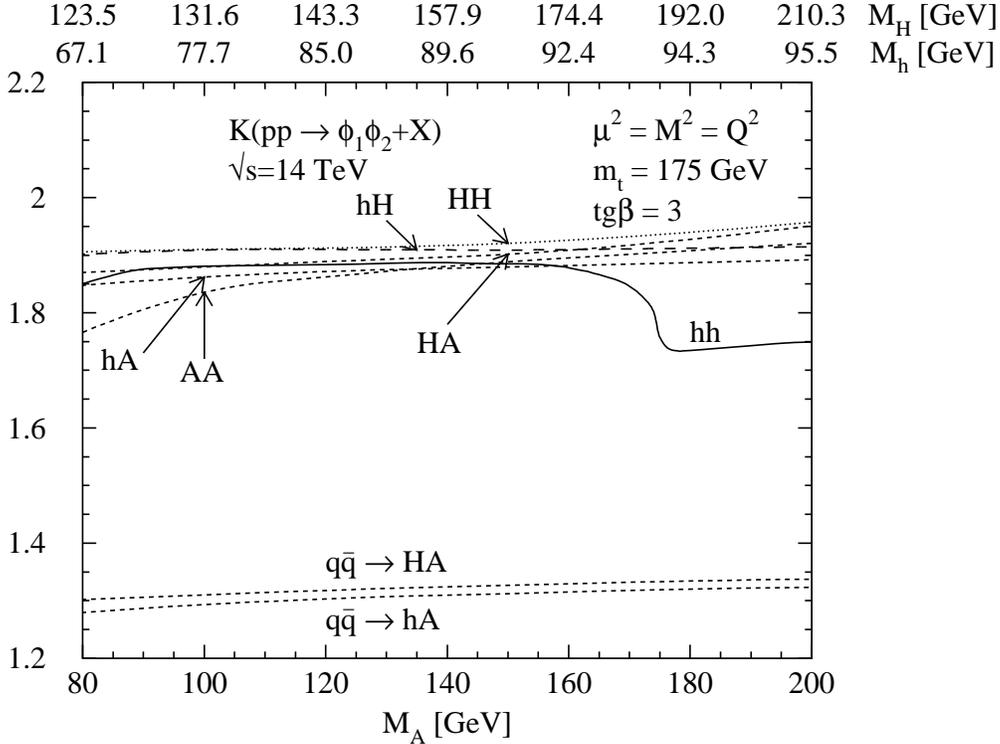}
\end{turn}
\vspace*{-0.5cm}
\caption[]{\it \label{fg:kmssm} K factors of the QCD-corrected gluon-fusion
and Drell--Yan like cross sections $\sigma(pp \to \phi_1\phi_2+X)$ at the LHC
with c.m.~energy $\sqrt{s}=14$~TeV. 
}
\end{figure}
The $K$ factors for all Higgs-boson 
pair-production processes for $\tgb=3$ are
presented in Fig.~\ref{fg:kmssm} as a function of the pseudoscalar Higgs mass
$M_A$. The numbers at the top are the corresponding values of the Higgs masses
$M_h$ and $M_H$. The total $K$ factors increase the Drell--Yan like cross
sections by about 30\% and the $gg$ production cross sections
by about 60--100\%, thus 
signalling the importance of the QCD corrections.
These are dominated by soft and collinear gluon radiation from
$gg$ and $q\bar q$ initial states, similar to the SM case discussed above in
detail. The
sharp decrease of the $K$ factor for $gg\to hh$ at $M_A\sim 175$ GeV originates
from the resonance contribution $gg\to H\to hh$, which is kinematically
forbidden below this mass and allowed above. The $K$ factor of resonant
single-Higgs production is smaller than the one of continuum $hh$
production.

\begin{figure}[p]
\vspace*{0.5cm}
\hspace*{0.5cm}
\begin{turn}{-90}%
\epsfxsize=8cm \epsfbox{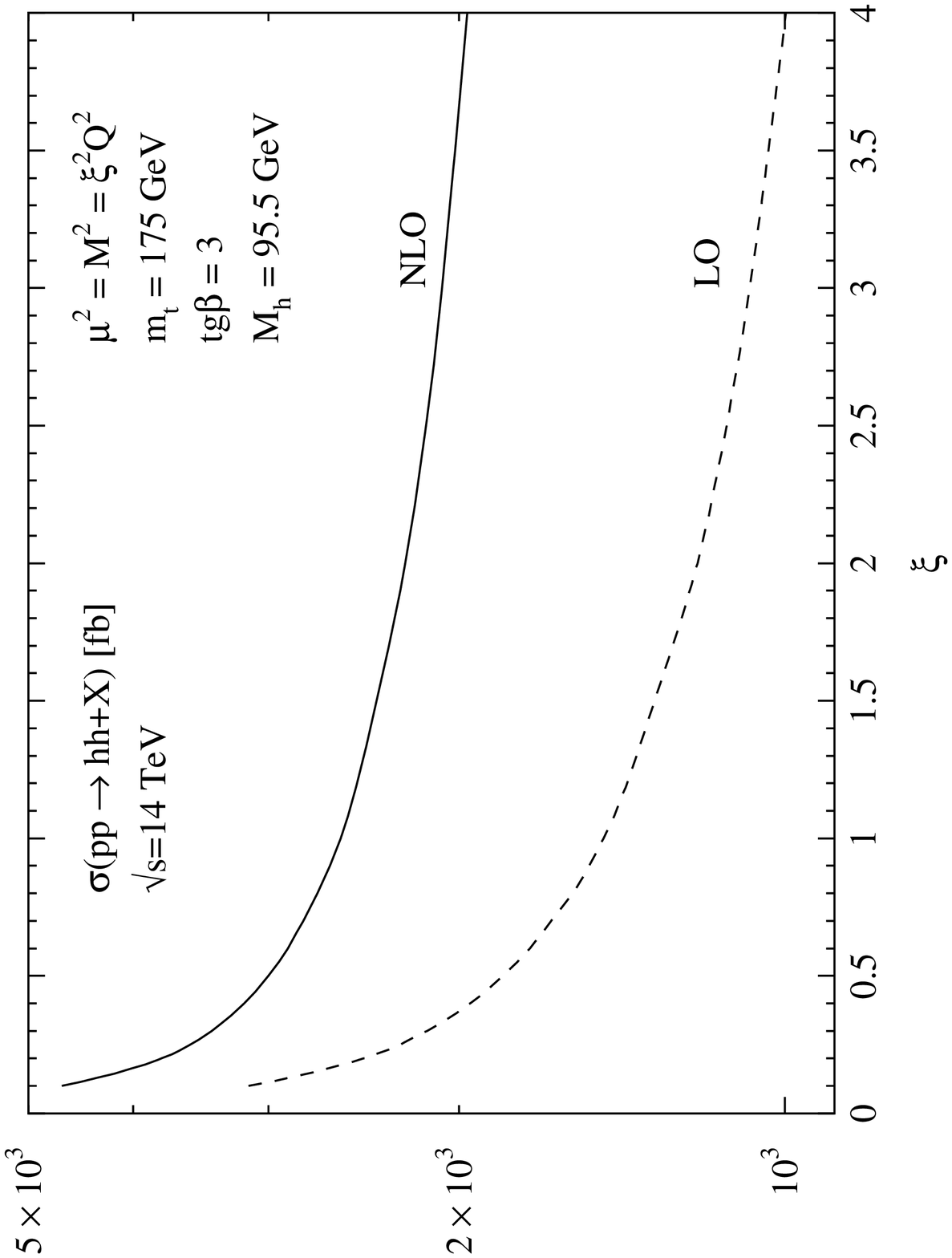}
\end{turn}
\vspace*{-0.3cm}
\caption[]{\it \label{fg:mssmscale} The renormalization and factorization scale
dependence of the Higgs pair-production cross section $\sigma(pp\to hh+X)$
at LO and NLO for $\tgb=3$ and a Higgs mass $M_h = 95.5$~GeV
($M_A=200$~GeV).}
\hspace*{0.0cm}
\begin{turn}{-90}%
\epsfxsize=10cm \epsfbox{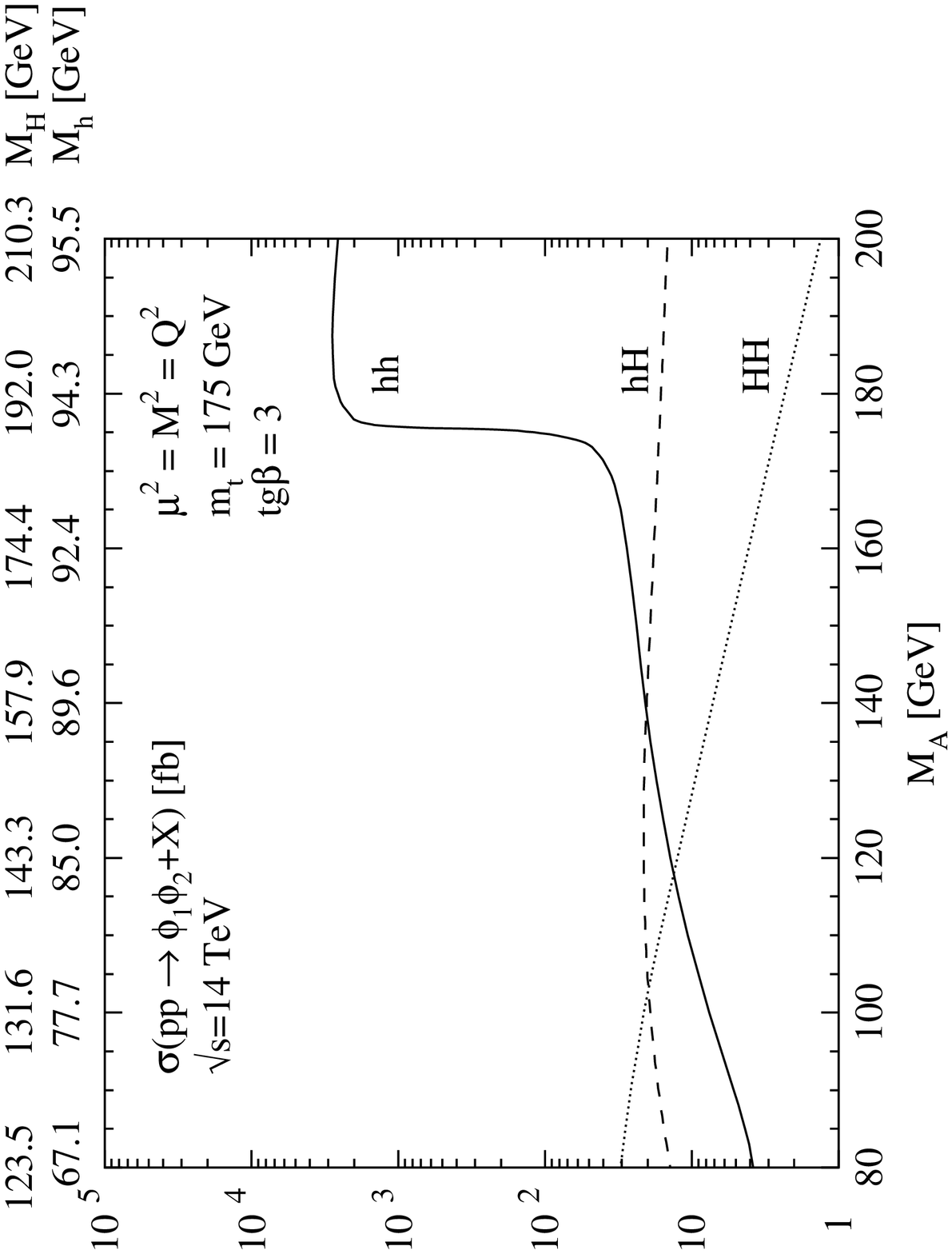}
\end{turn}
\vspace*{-0.5cm}
\caption[]{\it \label{fg:mssmcxn} Scalar Higgs-boson pair-production cross
sections $\sigma(pp\to hh,hH,HH + X)$ at NLO as functions of the pseudoscalar
Higgs mass $M_A$ for $\tgb=3$.}
\end{figure}

In spite of the large size of the NLO contributions, 
the scale dependence significantly decreases 
as can be inferred from Fig.~\ref{fg:mssmscale},
which presents the scale dependence of the cross section $\sigma(pp\to hh+X)$
at LO and NLO for a 
light-scalar-Higgs mass $M_h=95.5$~GeV, corresponding to $M_A=200$~GeV.
Thus, the QCD-corrected results turn out to be reliable within about 
$\pm 20\%$. 
The total NLO cross sections for the processes $gg\to hh$,
$hH$, $HH$ are presented in Fig.~\ref{fg:mssmcxn} as a function of the
pseudoscalar Higgs mass $M_A$. 
The sharp increase of the $hh$ cross section at $M_A\sim 175$~GeV is due to
the fact that the resonant $gg\to H\to hh$ process opens up above this mass
value, while it is kinematically forbidden below.
There are sizeable regions where the cross sections exceed a level of 10~fb.
This is in contrast to $AA$ production, the cross section of which is
smaller than 1~fb in the considered parameter space.

\begin{figure}[p]
\hspace*{0.0cm}
\begin{turn}{-90}%
\epsfxsize=10cm \epsfbox{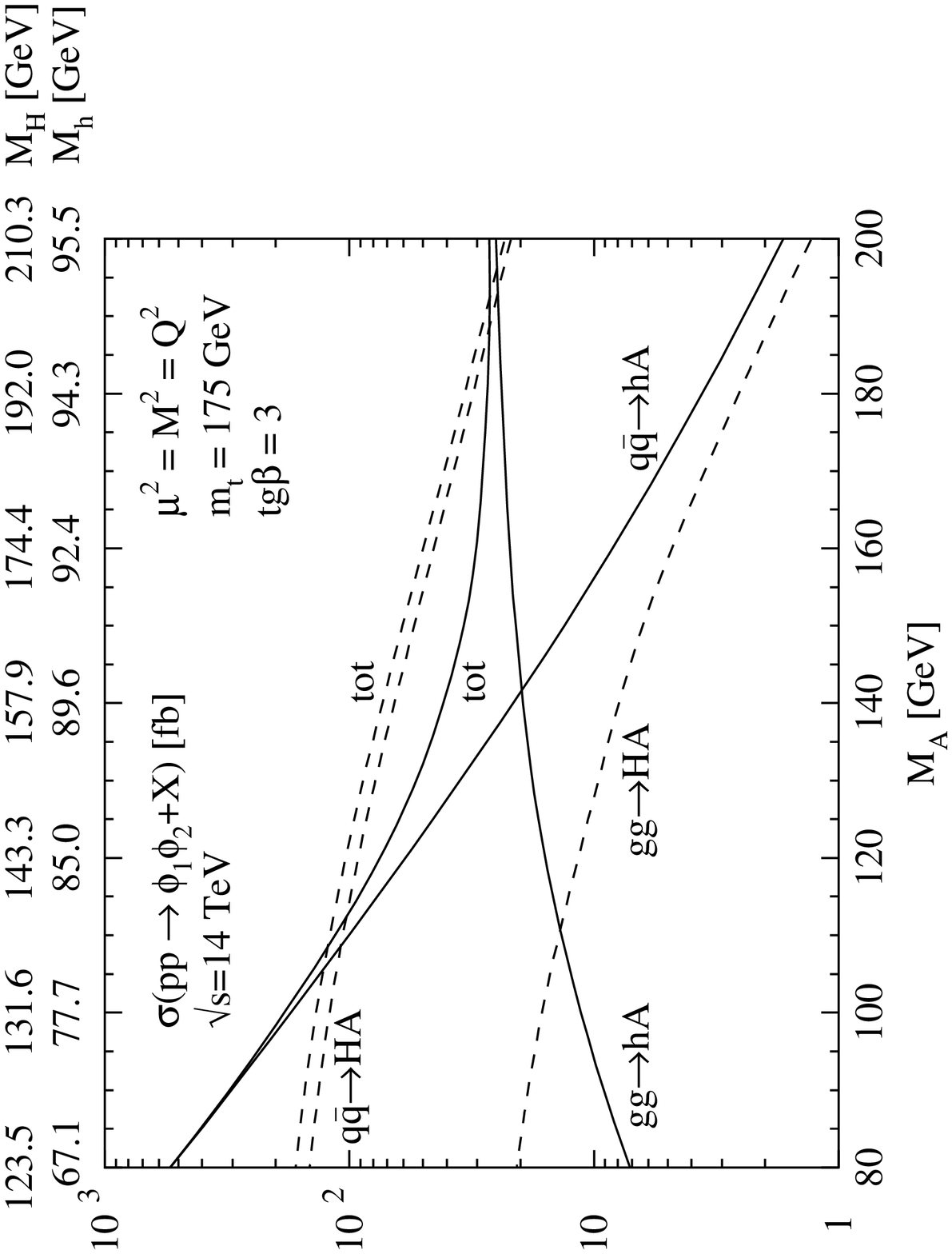}
\end{turn}
\vspace*{-0.5cm}
\caption[]{\it \label{fg:cxnha} Scalar and pseudoscalar
Higgs-boson pair-production cross sections $\sigma(pp\to hA,HA + X)$ at NLO
as functions of the pseudoscalar Higgs mass $M_A$ for $\tgb=3$ and their
individual contributions from $gg$ and $q\bar q$ collisions. The full lines
correspond to $hA$ and the broken ones to $HA$ production.}
\end{figure}

The total cross sections for the processes $pp\to hA,HA$ and their individual
contributions from the $gg$ and $q\bar q$ initiated processes are presented
in Fig.\ \ref{fg:cxnha} as a function of the pseudoscalar mass $M_A$ for
$\tgb=3$. 
While for 
$HA$-pair production gluon fusion $gg\to HA$ is always suppressed against
the Drell--Yan like process $q\bar q\to HA$, both corresponding processes are
competitive for the light scalar Higgs particle $h$. Especially for smaller
masses $M_A$ the total cross sections for $pp\to hA,HA$ exceed 10~fb and thus
potentially provide the possibility to detect these processes. Note, however,
that only the gluon-fusion processes are sensitive to the trilinear Higgs
couplings.

Finally, it should be noted, that in all neutral Higgs-pair processes the
residual theoretical uncertainties reduce to a level of 20\% after
including the QCD corrections.

\begin{figure}[t]
\hspace*{0.0cm}
\begin{turn}{-90}%
\epsfxsize=10cm \epsfbox{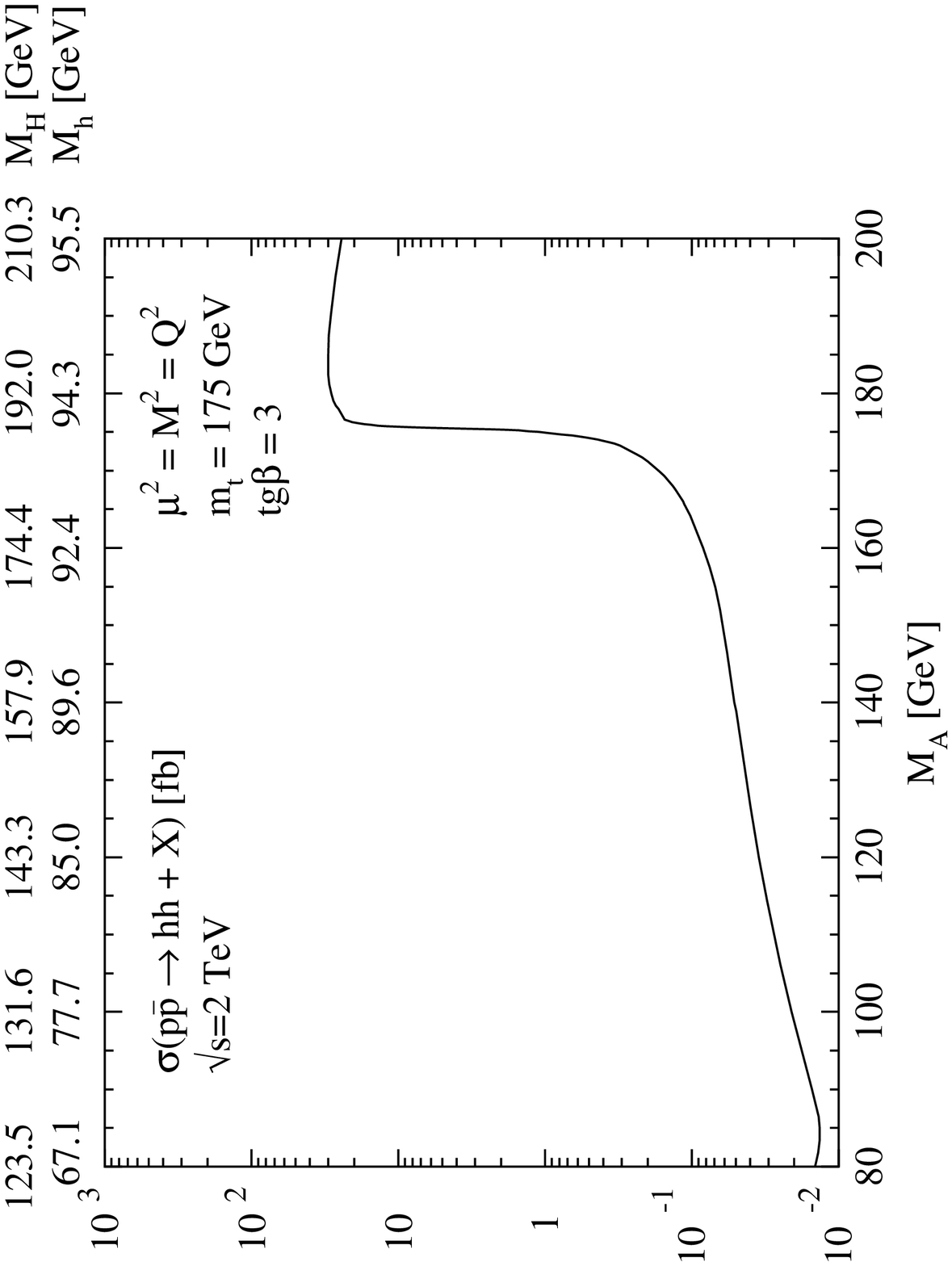}
\end{turn}
\vspace*{-0.5cm}
\caption[]{\it \label{fg:tevatron} Scalar Higgs-boson pair-production cross
section $\sigma(p\bar p\to hh + X)$ at NLO as a function of the pseudoscalar
Higgs mass $M_A$ for $\tgb=3$ at the Tevatron $p\bar p$ collider with c.m.\
energy $\sqrt{s}=2$ TeV.}
\end{figure}
The corresponding NLO cross section for $hh$ production at the Tevatron with
c.m.\ energy $\sqrt{s}=2$ TeV is presented in Fig.\ \ref{fg:tevatron} as a
function of the pseudoscalar mass $M_A$. For $M_A \gsim 175$ GeV it exceeds
10 fb, since the resonant $gg\to H\to hh$ is kinematically possible and
dominant in this mass region. The signatures of this process are
$b\bar b b\bar b$ and $b\bar b \tau^+ \tau^-$ final states.

Trilinear MSSM Higgs couplings can be measured at high energy
$e^+e^-$ colliders \cite{eehhmssm}, if sufficient numbers of signal events
for Higgs pair production will be produced. At the necessary high luminosities
background processes do not cause any problem.

\section{Conclusions}
\label{se:concl}
We have presented a complete calculation of the two-loop QCD corrections to
neutral-Higgs pair production at the LHC 
via gluon fusion in the limit of a heavy top quark. 
This approximation is at least reliable if the invariant mass of the produced
Higgs-boson pair is below the $tt$ threshold of the mediating top-quark
loops.
We have analyzed the results within the SM and the MSSM
and found large increases of the cross sections by about 60--100\%.
The QCD corrections to the associated production processes $q\bar q\to hA,HA$
coincide with those to the Drell--Yan process $q\bar q \to Z^*$, thus
increasing the cross sections by about 30\%.
The QCD corrections stabilize the theoretical predictions compared 
with the LO results, which exhibit large theoretical uncertainties. After
including the QCD corrections, the remaining theoretical uncertainties are
reduced to a level of about 20\%.

Except for $AA$-pair production, all Higgs-pair production cross
sections of the SM and the MSSM exceed 10~fb at LHC energies in certain 
regions of the parameter spaces. As soon as the $H\to hh$ channel opens,
the cross section for $pp\to hh+X$ reaches even $10^3$~fb at the LHC and
is still larger than 10~fb at the Tevatron for $\tgb=3$.

Moreover, we have shown that the NLO prediction of Higgs-boson pair
production is sensitive to a deviation of the trilinear Higgs coupling
from its SM value, rendering this process well suited for studying this
coupling in $pp$ collisions. \\[1.0cm]

\noindent
{\bf Acknowledgements} \\[0.5cm]
We would like to thank T.\ Plehn and P.M.\ Zerwas for useful discussions.
Moreover, we are grateful to P.M.\ Zerwas for reading the manuscript and his
valuable comments.


\begin{thebibliography}{99}

\bibitem{higgs} P.\ W.\ Higgs, 
Phys.\ Lett.\ {\bf 12} (1964) 132 and
Phys.\ Rev.\ {\bf 145} (1966) 1156; \\
F.\ Englert and R.\ Brout, Phys.\ Rev.\ Lett.\ {\bf 13} (1964) 321; \\
G.\ S. Guralnik, C.\ R.\ Hagen and T.\ W.\ Kibble, Phys.\ Rev.\ Lett.\ {\bf 13}
(1964) 585.

\bibitem{hunter} For reviews 
on the Higgs sector in the Standard Model and in
its supersymmetric extensions, see J.F.\ Gunion, H.E.\ Haber, G.L.\ Kane and
S.\ Dawson, 
{\it The Higgs Hunter's Guide} (Addison--Wesley, Reading, Mass., 1990); \\
S.\ Dawson, preprint BNL-HET-SD-97-004, to be published in 
{\it Perspectives in Higgs Physics}, G. Kane, ed., hep-ph/9703387; \\
M.\ Spira and P.M.\ Zerwas, preprint CERN-TH/97-379, Lectures at 
{\it 36th
Internationale Universit\"atswochen f\"ur Kernphysik und Teilchenphysik},
Schladming 1997, hep-ph/9803257.

\bibitem{mssmrad1} 
Y.\ Okada, M.\ Yamaguchi and T.\ Yanagida, Prog.\ Theor.\ Phys.\ {\bf 85}
(1991) 1; \\
H.E.\ Haber and R.\ Hempfling, Phys.\ Rev.\ Lett.\ {\bf 66} (1991) 1815; \\
J.\ Ellis, G.\ Ridolfi and F.\ Zwirner, Phys.\ Lett.\ {\bf B257} (1991) 83.

\bibitem{mssmrad2a} 
R.\ Hempfling and A.\ Hoang, Phys.\ Lett.\ {\bf B331} (1994) 99; \\
S.\ Heinemeyer, W.\ Hollik and G.\ Weiglein, preprint KA-TP-2-1998,
hep-ph/9803277.

\bibitem{mssmrad2} 
M.\ Carena, J.R.\ Espinosa, M.\ Quir\'os and C.E.M.\ Wagner, Phys.\ Lett.\ {\bf
B355} (1995) 209; \\
M.\ Carena, M.\ Quir\'os and C.E.M.\ Wagner, Nucl.\ Phys.\ {\bf B461} (1996)
407.

\bibitem{ppvhh} V.\ Barger, T.\ Han and R.J.N.\ Phillips, 
Phys.\ Rev.\ {\bf D38} (1988) 2766.

\bibitem{ppvvhh} D.A.\ Dicus, K.J.\ Kallianpur and S.S.D.\ Willenbrock, 
Phys.\ Lett.\ {\bf B200} (1988) 187; \\
A.\ Abbasabadi, W.W.\ Repko, D.A.\ Dicus and R.\ Vega, 
Phys.\ Rev.\ {\bf D38} (1988) 2770; Phys.\ Lett.\ {\bf B213} (1988) 386.

\bibitem{ppqqhh} K.J.\ Kallianpur, Phys.\ Lett.\ {\bf B215} (1988) 392.

\bibitem{gghhsm} E.W.N.\ Glover and J.J.\ van der Bij, Nucl.\ Phys.\ {\bf B309}
(1988) 282.

\bibitem{gghhlo} T.\ Plehn, M.\ Spira and P.M.\ Zerwas, Nucl.\ Phys.\
{\bf B479} (1996) 46 and Erratum.

\bibitem{djlhc} A. Djouadi, talk at the Workshop 
{\it Theory of LHC Processes},
CERN, Geneva, 1998.

\bibitem{soft} M.\ Kr\"amer, E.\ Laenen and M.\ Spira, Nucl.\ Phys.\ {\bf B511}
(1998) 523.

\bibitem{gghlim} A.\ Djouadi, M.\ Spira and P.M.\ Zerwas, Phys.\ Lett.\
{\bf B264} (1991) 440; \\
S.\ Dawson, Nucl.\ Phys.\ {\bf B359} (1991) 283.

\bibitem{higgsqcd} M.\ Spira, A.\ Djouadi, D.\ Graudenz and P.M.\ Zerwas,
Nucl.\ Phys.\ {\bf B453} (1995) 17.

\bibitem{let0}J.\ Ellis, M.K.\ Gaillard and D.V.\ Nanopoulos, Nucl.\ Phys.\
{\bf B106} (1976) 292; \\
A.I.\ Vainshtein, M.B.\ Voloshin, V.I.\ Zakharov and M.A.\ Shifman, Sov.\ J.\
Nucl.\ Phys.\ {\bf 30} (1979) 711.

\bibitem{let1} B.A.\ Kniehl and M.\ Spira, Z.\ Phys.\ {\bf C69} (1995) 77.

\bibitem{abj} S.L.\ Adler, Phys.\ Rev.\ {\bf 177} (1969) 2426; \\
J.S.\ Bell and R.\ Jackiw, Nuovo Cimento {\bf 60A} (1969) 47.

\bibitem{agg} A.\ Djouadi, M.\ Spira and P.M.\ Zerwas, Phys.\ Lett.\ {\bf B311}
(1993) 255; \\
M.\ Spira, A.\ Djouadi, D.\ Graudenz and P.M.\ Zerwas, Phys.\
Lett.\ {\bf B318} (1993) 347; \\
R.P.\ Kauffman and W.\ Schaffer, Phys.\ Rev.\ {\bf D49} (1994) 551.

\bibitem{abt} S.L.\ Adler and W.A.\ Bardeen, Phys.\ Rev.\ {\bf 182} (1969)
1517; \\
R.\ Jackiw, 
{\it Lectures on Current Algebra and its Applications} (Princeton
University Press, 1972).

\bibitem{altpar} G.\ Altarelli and G.\ Parisi, Nucl.\ Phys.\ {\bf B126}
(1977) 298.

\bibitem{FA} J.~K\"ublbeck, M.~B\"ohm and A.~Denner, 
Comput.\ Phys.\ Commun.\ {\bf 60} (1990) 165; \\
H.~Eck and J.~K\"ublbeck, {\it Guide to FeynArts 1.0\/},
University of W\"urzburg, 1992.

\bibitem{sm97}
V.A.~Smirnov, Phys.\ Lett.\ {\bf B394} (1997) 205.

\bibitem{asymp}
S.G.~Gorishny, Nucl.\ Phys.\ {\bf B319} (1989) 633; \\
V.A.~Smirnov, Commun.\ Math.\ Phys.\ {\bf 134} (1990) 109; 
{\it Renormalization and asymptotic expansions} 
(Birkh\"auser Verlag, Basel, 1991); \\
F.V.~Tkachov, Int.\ J.\ Mod.\ Phys.\ {\bf A8} (1993) 2047; \\
G.B.~Pivovarov and F.V.~Tkachov, Int.\ J.\ Mod.\ Phys.\ {\bf A8} (1993) 2241. 

\bibitem{math} S.~Wolfram, {\it Mathematica --- A System for Doing
Mathematics by Computer} (Addison-Wesley, Redwood City, CA, 1988).

\bibitem{dimreg}
G.~'t~Hooft and M.~Veltman, Nucl.\ Phys.\ {\bf B44} (1972) 189; \\
P.\ Breitenlohner and D.\ Maison, Commun.\ Math.\ Phys.\ {\bf 52} (1977) 11.

\bibitem{dyqcd} W.\ Furmanski and R.\ Petronzio, Z.\ Phys.\ {\bf C11} (1982)
293 and references therein.

\bibitem{cteq4} H.L.\ Lai, J.\ Huston, S.\ Kuhlmann, F.\ Olness, J.\
Owens, D.\ Soper, W.K.\ Tung and H.\ Weerts, Phys.\ Rev.\ {\bf D55}
(1997) 1280. 

\bibitem{eehhsm} G.J.\ Gounaris, D.\ Schildknecht and F.M.\ Renard, 
Phys.\ Lett.\ {\bf B83} (1979) 191; (E) {\bf B89} (1980) 437; \\
J.F.\ Gunion, L.\ Roszkowski, A.\ Turski, H.E.\ Haber, G.\ Gamberini, B.\
Kayser, S.F.\ Novaes, F.\ Olness and J.\ Wudka, Phys.\ Rev.\ {\bf D38} (1988)
3444; \\
V.\ Barger and T.\ Han, Mod.\ Phys.\ Lett.\ {\bf A5} (1990) 667; \\
F.\ Boudjema and E.\ Chopin, Z.\ Phys.\ {\bf C73} (1996) 85.

\bibitem{richter} E.\ Richter-Was, D.\ Froidevaux, F.\ Gianotti, L.\ Poggioli,
D.\ Cavalli and S.\ Resconi, 
preprint CERN-TH/96-111.

\bibitem{higgsrev} For recent reviews, 
see e.g.\ A.\ Djouadi, Int.\ J.\ Mod.\
Phys.\ {\bf A10} (1995) 1; \\
M.\ Spira, Fortschr.\ Phys.\ {\bf 46} (1998) 203.

\bibitem{eehhmssm} A.\ Djouadi, H.E.\ Haber and P.M.\ Zerwas, Phys.\ Lett.\
{\bf B375} (1996) 203.

\end{thebibliography}
\end{document}